# Deep learning enhanced individual nuclear-spin detection

Kyunghoon Jung[1]*, M.H. Abobeih[2,3]*, Jiwon Yun[1], Gyeonghun Kim[1], Hyunseok Oh[1], Henry Ang[4], T.H. Taminiau[2,3]**, and Dohun Kim[1]**

[1]Department of Physics and Astronomy, and Institute of Applied Physics, Seoul National University, Seoul 08826, Korea

[2]QuTech, Delft University of Technology, PO Box 5046, 2600 GA Delft, The Netherlands

[3]Kavli Institute of Nanoscience Delft, Delft University of Technology, PO Box 5046, 2600 GA Delft, The Netherlands

[4]Department of Physics and Astronomy, University College London, Gower Street, London, WC1E 6BT United Kingdom

* These authors contributed equally to this work.

**Corresponding author: t.h.taminiau@tudelft.nl, dohunkim@snu.ac.kr

**The detection of nuclear spins using individual electron spins has enabled new opportunities in quantum sensing and quantum information processing. Proof-of-principle experiments have demonstrated atomic-scale imaging of nuclear-spin samples and controlled multi-qubit registers. However, to image more complex samples and to realize larger-scale quantum processors, computerized methods that efficiently and automatically characterize spin systems are required. Here, we realize a deep learning model for automatic identification of nuclear spins using the electron spin of single nitrogen-vacancy (NV) centers in diamond as a sensor. Based on neural network algorithms, we develop noise recovery procedures and training sequences for highly non-linear spectra. We apply these methods to experimentally demonstrate fast identification of 31 nuclear spins around a single NV center and accurately determine the hyperfine parameters. Our methods can be extended to larger spin systems and are applicable to a wide range of electron-nuclear interaction strengths. These results enable efficient imaging of complex spin samples and automatic characterization of large spin-qubit registers.**

Recent advances in the control of single electron spins associated to defects in solids have enabled the sensing, imaging and control of individual nuclear spins[1-16]. From a quantum sensing perspective, this has enabled the detection and imaging of nuclear spins with atomic-scale resolution and single spin sensitivity, in systems of up to 27 spins[5,10,17-20]. From a quantum information perspective, controlling individual nuclear spins provides quantum registers for quantum computation and optically-connected quantum networks[15,21-24]. Proof-of-principle experiments have demonstrated quantum registers with 10+ qubits[15,21-25], elementary quantum algorithms and error correction protocols[6,26-31], and key quantum network protocols such as entanglement distillation[32,33].

An important task in both these application fields is to detect and identify the nuclear spins, and to characterize the electron-nuclear interaction. For imaging larger, more complex, spin structures and for the realization of large-scale quantum networks that consist of many multi-qubit devices, it is required to develop objective and automated methods that can efficiently identify signatures of nuclear spins and determine coupling parameters from experimental spectroscopy.

In this work, we develop neural-network-based algorithms that can efficiently and automatically detect nuclear spins by their coupling to a single electron spin. We focus on CPMG-type dynamical decoupling spectroscopy[2-5,22,34,35], which is widely employed for single nuclear spin detection and control[3,36] and is a common starting point for more advanced spectroscopy methods[10,11,14,19]. While our methods are general, we exemplify them through experiments on a single nitrogen-vacancy (NV) center in diamond with nearby naturally abundant $^{13}$C nuclear spins[3,6,22]. We show that our deep learning approach enables fast automatic nuclear spin detection and hyperfine parameter estimation for 31 individual spins.

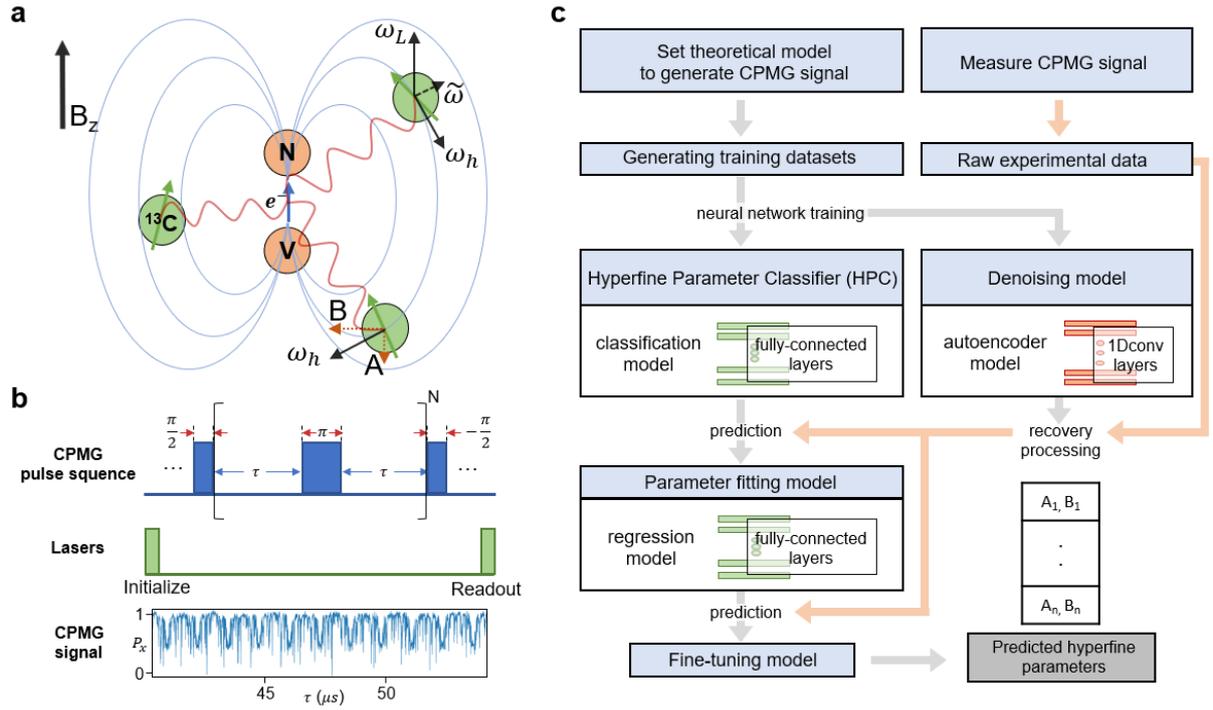

**Figure. 1 General procedure for identifying hyperfine parameters of $^{13}$C nuclear spins. a.** Schematic diagram showing the configuration of an electron spin within the nitrogen-vacancy (NV) center magnetic dipole field (blue oval curves) and $^{13}C$ nuclear spins (green circles) interacting with the NV center via hyperfine interaction. $B_z$ is the external magnetic field, $\omega_L$ is the Larmor frequency, $\omega_h = \sqrt{A^2 + B^2}$, and $\tilde{\omega} = \sqrt{(A+\omega_L)^2 + B^2}$ where $A$ ($B$) is the longitudinal (transverse) hyperfine interaction parameter. **b.** Typical dynamical decoupling pulse sequence (Carr-Purcell-Meiboom-Gill, CPMG) used for experimental nuclear spectroscopy. The bottom panel shows an example of experimental CPMG data from which electron-nuclear hyperfine interaction is analyzed. **c.** Pseudo-algorithm for training and hyperfine-parameters-prediction sequences including hyperfine parameter classifier (HPC), denoise and signal recovery, regression-based fitting, and fine-tuning models. The flow of experimental processes (computational processes) is on the red (gray) arrows.

**Results**

**Theoretical model and overall procedure.**

Figure 1a shows a schematic of the electron-nuclear spin complex considered in this work. The NV center, an impurity in the diamond crystal lattice, acts as a sensitive probe for the surrounding nuclear-spin environment. The ground state electron spin of the NV center can be initialized and measured using spin-dependent fluorescence and can be manipulated by microwaves[37]. In typical dynamical decoupling spectroscopy, for example based on Carr-Purcell-Meiboom-Gill (CPMG) pulse sequence[22] shown in Fig 1b, the interaction of the electron with its nuclear spin environment leads to sudden and periodic losses of coherence at specific pulse timings. The magnitude and position of the dip in coherence depends on the longitudinal (transverse) hyperfine coupling parameter $A$ ($B$). The CPMG signal is given by the probability $P_x$ that the NV center's spin state is preserved. In the absence of nuclear-nuclear interactions this can be described as[3],

$$P_x = \frac{1 + \prod_{k=1}^{n} M_k}{2} \tag{1}$$

$$M_k = 1 - m_{k,z}^2 \frac{(1-\cos\alpha_k)(1-\cos\beta)}{1+\cos\alpha_k \cos\beta - m_{k,z}\sin\alpha_k \sin\beta} \sin^2 \frac{N\phi_k}{2} \tag{2}$$

$$\cos\phi_k = \cos\alpha_k \cos\beta - m_{k,z}\sin\alpha_k \sin\beta \tag{3}$$

, where $m_{k,z} = (A_k + \omega_L)/\tilde{\omega}_k$, $m_{k,x} = B_k/\tilde{\omega}$, $\tilde{\omega}_k = \sqrt{(A_k+\omega_L)^2 + B_k^2}$, $\alpha_k = \tilde{\omega}_k\tau$, $\beta = \omega_L\tau$, $\tau$ is half of the delay between $\pi$ pulses, $k$ indicates $k^{\text{th}}$ nuclear spin, $n$ is the total number of nuclear spins, $\omega_L$ is the Larmor frequency, and $N$ is the repetition number of the unit CPMG pulse (see Fig. 1b). The CPMG signal is given by the multiplication of all the $M_k$'s for $n$ nuclear spins as depicted in Eqn (1) and (2). This characteristic introduces an additional complexity compared to conventional nuclear magnetic resonance (NMR) signals[38-40] and, along with the decoherence and environmental noises, makes existing NMR peak decomposition packages[41-46] ineffective for analyzing the signal.

The main task of our deep learning model is to efficiently encode the features of each $k^{\text{th}}$

nuclear spin in Eqn. (1). Once successfully trained, the models can determine $A_k$ and $B_k$ of each nuclear spin from the experimental spectroscopy data (see the bottom panel of Fig. 1b for an example). Figure 1c shows the overall procedure to achieve this task. First, the measurements of CPMG signals and the implementations for generating datasets and training deep learning models are conducted simultaneously. Generating datasets for both HPC models and denoising models is performed using the theoretical model in Eqn. (1). Second, via the generated training datasets, denoising models are trained to reduce noise and HPC models are trained to identify whether specific hyperfine parameters exist in the data or not. Third, to enhance signal-to-noise ratio, the raw noisy CPMG signal is pre-processed by the trained denoising model and decoherence recovery process and is fed into the trained HPC models. Fourth, using the outputs of the HPC models, an additional deep learning-based regression model is adapted to further restrict possible hyperfine parameter combinations. Lastly, in the auto fine-tuning phase, the prediction of the regression model is used as initial values of the hyperfine parameters, and automatic numerical fitting is performed.

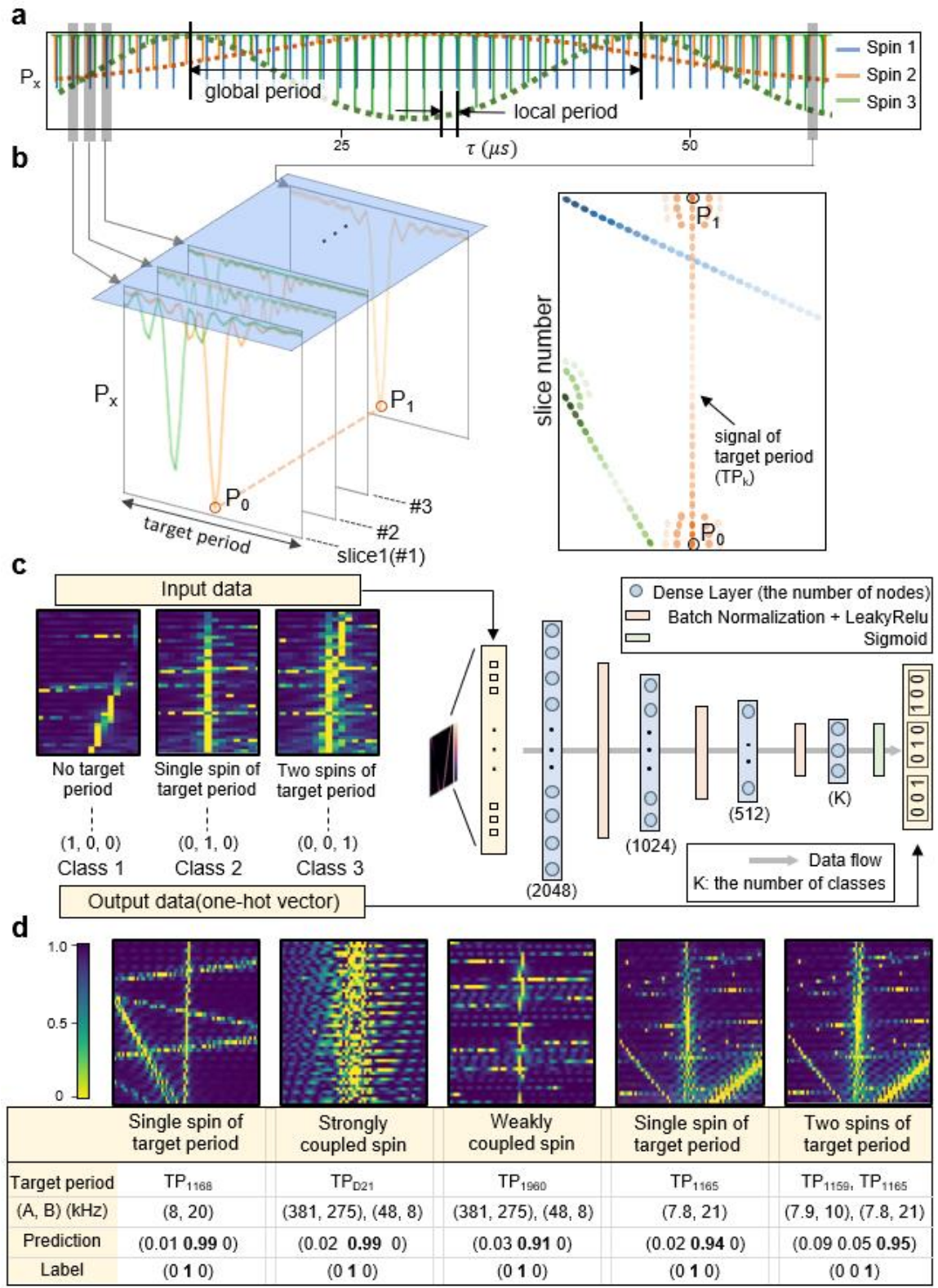

**Figure. 2. Individual spin signature identification by hyperfine parameter classifier (HPC) deep learning model. a.** Simulated CPMG signal with three spins of different (*A*, *B*) values showing general

features of nuclear spectra including local and global periods (see text for definitions). **b.** Concept of the data conversion into 2D images by slicing and stacking the data fragments with specified target period $TP_k$. This reveals the signature of a target signal as a vertical line with vanishing slope, which is generally superposed with other interfering nuclear spin signals. **c.** Training datasets and architecture of the HPC model in a case of classifying three classes (K=3), where K is the number of nodes of the last layer. The input data consists of three different classes, where each class corresponds to the number of existing nuclear spins with the target period, and the output data is *one-hot vector* form assigned to each class. For example, Class 1 (Class 2) means that no (one) spin with the target period exists. **d.** Example predictions of the HPC model depending on hyperfine coupling strength (first to third panels) and proximity to similar period (fourth and fifth panels). For all cases, the HPC model predicts correct spin signatures corresponding to input signals showing good consistency between the predicted vectors and the output vectors. (for example, in the first panel, the predicted vector is (0.01 0.99 0) and the output vector is (0 1 0)). The color scale bar in all 2D images ranges from 0 to 1.

**Data representation for nuclear spin detection.**

The qualitative features of a typical dynamical decoupling signal are as follows. First, the coherence dip of $k^{th}$ nuclear spin is periodic with approximate periodicity[3] (local period)

$$TP_k = 2\pi / (\tilde{\omega}_k + \omega_L) \tag{4}$$

where $\tilde{\omega}_k = \sqrt{(A_k + \omega_L)^2 + B_k^2}$, and $A_k$ and $B_k$ are hyperfine parameters of the '$k^{th}$ target' local period (see Supplementary Fig. 1 for the range of $A$ and $B$ values used in the proposed model). Second, the envelop of the coherence dip amplitudes as a function of $\tau$ is periodic essentially showing periodic quantum entanglement evolution with the resonant nuclear spin[8] (global period). Third, each coherence dip can show additional fringes depending on the hyperfine interaction strength. In the strong coupling regime, for example when $B > 100$ kHz, the CPMG signal can exhibit multiple and large fringe oscillations[3,6] (Eqn. 2,3). While conventional numerical peak detection or Fourier transform analysis is inefficient in the presence of these oscillating signals, below we show that the deep learning approach offers an excellent alternative route to solve the problem.

In principle, supervised learning algorithms can be applied using the theoretical model given by Eqns. (1)-(3) for this nominally multi class classification problem[47-51]. The data preparation and training, however, is challenging in that, (1) the number of nuclear spins interacting with central NV center is not known *a priori* and (2) the number of possible (*A*, *B*) pair combinations for a given number of surrounding nuclear spins is large. Brute force generation of large datasets with variable number of nuclear spins is impractical and generally not reliable to represent possible spin configurations unambiguously.

We convert the multi class classification problem to that of single class by reorganizing the data so that the deep learning model focuses on identifying a single target spin. Figure 2b shows the general concept of this conversion. By cutting the CPMG signal according to the $TP_k$ of a target spin and making a 2D image by stacking multiple slices, the difference of the local periods between two spins can be distinguished. The features of the global period can be also analyzed by the distribution of pixel values on the vertical axis. With this representation, the deep learning model analyzes whether the target spin signal marked by a vertical line exists in the 2D image. Moreover, non-linear oscillations near the main coherence dip in the strong coupling regime, which are difficult to address by hand-crafted coding, generally appear as fringe patterns in this data representation. The deep learning model shows a strong ability of classifying target signals in the presence of these interfering patterns through image recognition[52,53].

**Deep learning model for classification.**

Focusing on the local period of a specific target spin signal, we develop a set of deep learning models, coined HPC, each of which classifies the existence of a specific period of hyperfine-induced coherence dips in the data. Figure 2c illustrates a structure of the HPC model and training datasets by exemplifying a case of classifying three different classes (see a detailed implementation of generating training datasets in Supplementary Note 1). The input training data is prepared along with three output classes as shown in Fig. 2c. Class 1 corresponds to data that does not contain a spin with the target period, class 2 is for one spin with the target period existing in the data, and class 3 for two spins with slightly dissimilar target periods in the data. The output data is denoted in *one-hot vector* form; (1, 0, 0), (0, 1, 0), and (0, 0, 1) corresponding to no, single, and double target periods, respectively. The model is trained to estimate the confidence score of each element of the three-dimensional vector according to the input image. The model consists of stacked layers of Dense layer, Batch Normalization layer[54] and LeakyRelu activation function as shown in Fig. 2c. The detailed procedure of the neural network development is described in the methods section, Supplementary Fig. 2, and Supplementary Note 1.

Panels in Fig. 2d show the classification results using our HPC model. The first panel is for the typical case that a single target period exists without strong disturbance from other spins nor spin bath signal, and the model successfully outputs a vector close to (0, 1, 0). The second panel shows the performance of the model for a strongly coupled single target spin where the parameters for spin number $TP_{D21}$ in Supplementary Table 1, taken from existing density functional theory (DFT) calculations[37], is used as an example. As mentioned above, although the spin signal is superposed with wide fringe patterns and oscillations, the model successfully identifies the signature of the target period with the output vector reaching (0.002, 0.99, 0). The third panel shows that the model also successfully classifies the target period even in the presence of other superposed strongly coupled spin signals. Furthermore, the fourth and the fifth panels give an example of the performance for input datasets with a single spin, $(A_2, B_2)/2\pi =(7.8, 20)$ (kHz) (fourth panel) and with two spins of similar local period, $(A_1, B_1)/2\pi =(7.9, 10)$, $(A_2, B_2)/2\pi =(7.8, 20)$ (kHz), (fifth panel). The model successfully distinguishes each case, showing high selectivity of the nuclear spins. Therefore, these results show that our deep learning model provides a promising approach to detect individual nuclear spins with high precision, high selectivity, and for a wide range of hyperfine strengths.

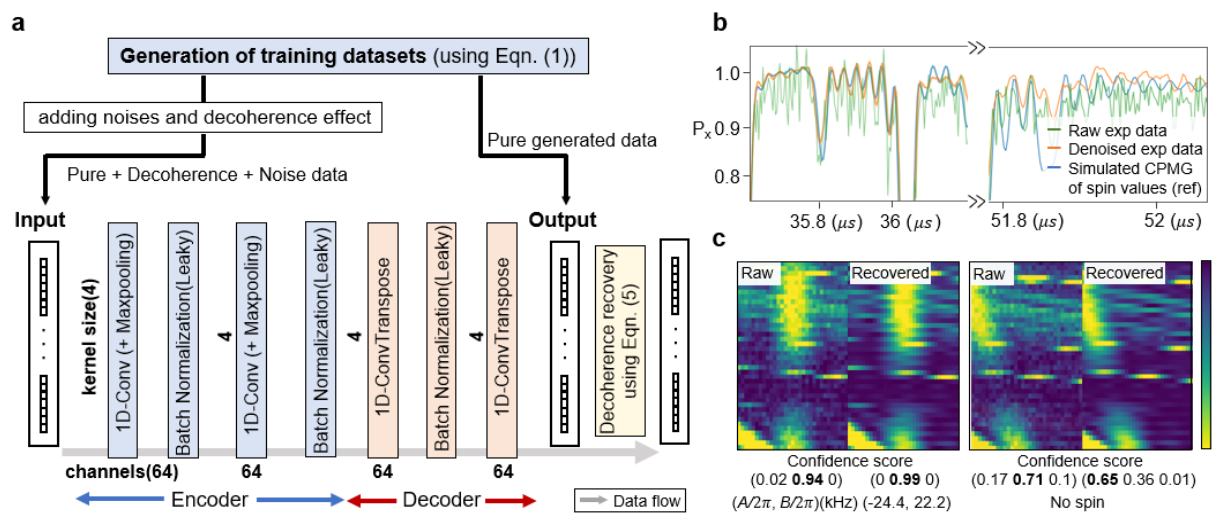

**Figure 3. Denoising and decoherence effect recovery procedure. a.** Architecture of the signal recovery model. The pure data is generated by Eqn. (1). The noisy data is generated with adding decoherence effects and noise to the pure data using Eqn. (5). The model is trained to reproduce denoised data from the noisy data and the decoherence effect is recovered using Eqn. (5). **b.** Raw experimental (green), pure (blue), and recovered (orange) CPMG data showing successful recovery of the fringe patterns in the presence of noise with a comparable amplitude. The figure also shows signal recovery performance in the long evolution time regime. **c.** The comparison of the raw experimental data with the noise recovered data in the 2D image form used for HPC model analysis showing an enhancement in signal-to-noise ratio and predictability. The color scale bar in all 2D images ranges from 0 to 1.

**Noise removal and decoherence effect recovery.**

Before evaluating the experimental CPMG signal by trained HPC models, we first pre-process the raw experimental data by a denoising model. Figure 3a shows the overall procedure. For the noise removal process, Gaussian noise with the standard deviation $\sigma = 0.05$ reflecting the experimental noise is added to the training datasets (see Supplementary Fig. 3). The decoherence effect is modelled by the approximate equation[3],

$$P_x = \frac{1}{2} M \cdot \exp(-\frac{\tau}{T})^n + \frac{1}{2} \tag{5}$$

, where $T$ accounts for dephasing of the electron spin, $n$ is an exponential power obtained by fitting the experimental data and $\tau$ is half of the inter-pulse delay. We use an autoencoder structure[55,56], which is an established structure to learn the representations of input data, to encode the features of the noisy input data and generate the denoised data. A one-dimensional convolution (1D CNN) layer[57], which is widely used to capture the features of one-dimensional data such as time-series signal, is employed for building the denoising neural network.

As shown in Fig. 3b, the signal recovery model effectively removes the noises while retaining nuclear spin signatures of the experimental data. This is highlighted with the capability of recovering detailed oscillatory features of the data where the amplitudes of signals are almost equivalent to the fluctuations due to noise. Panels in Fig. 3c compares the visibility of the spin signal of the raw (left panel) and the processed (right panel) data showing effective removal of experimental noise and enhancement of signal-to-noise ratio, leading to higher performance of prediction by the HPC model. After denoising the raw experimental data, the decoherence effect is recovered by applying Eqn. (5) to the denoised data. We find that the confidence scores from HPC evaluating denoised experimental data are, in general, a few percent higher than evaluating raw data (compare (0.02, 0.91, 0) vs. (0, 0.99, 0) in Fig. 3c) and in some cases false predictions of raw data are corrected in denoised data (compare (0.17 0.71 0.1) vs (0.65 0.36 0.01) in Fig. 3c), successfully showing the efficiency of our pre-processing model.

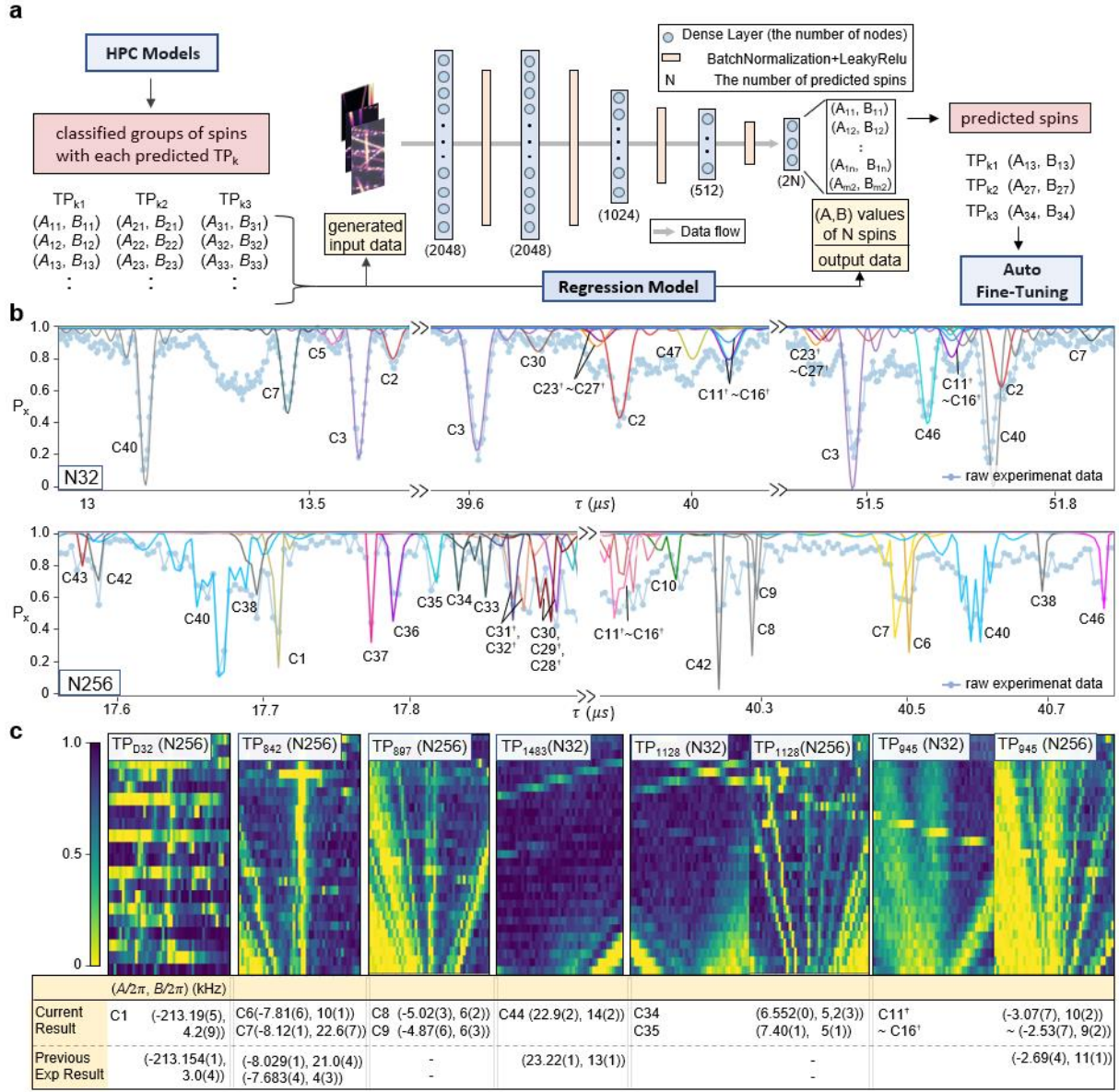

**Figure 4. Multiple nuclear spin detection from experimental data. a.** Procedures for tuning hyperfine parameters from predicted periods by HPC models. Training datasets of the regression model are generated using ($A$, $B$) pairs with the predicted period by HPC models. The regression model infers the single ($A$, $B$) pair that best fits with the features of the experimental data (reorganized as 2D image) including coherence dip amplitudes, envelop function, and fringe patterns. **b.** Multiple nuclear spin detection from the experimental CPMG data for $N$ =32 (top panel) and $N$ = 256 (bottom panel) using the same NV center. The panels show superimposed reproduced CPMG signal (solid curves) and the experimental data (dotted curve). The spin numbers (C#) indicated in the figure corresponds to the full

spin list summarized in Supplementary Table 2. **c.** Confirmation of detected hyperfine parameters for spins with large number of interfering signatures (1$^{st}$ panel), similar target periods (2$^{nd}$ and 3$^{rd}$ panels), weak local period signature (4$^{th}$ panel), small transverse hyperfine coupling (5$^{th}$ panel), and small longitudinal hyperfine coupling (6$^{th}$ panel). We compare the obtained values to the results reported in Ref. [19] (bottom row, see main text). The panels also show examples of spins with small A that were not detected in Ref. [19] (3rd, 5th, and 6th panels). The uncertainty in the last digit is given in parentheses. The color scale bar in all 2D images ranges from 0 to 1.

**Regression-based fitting, auto fine-tuning model, and application to experimental data.**

We now discuss the final stage of the deep learning protocol and the application of the overall procedure to experimental dynamical-decoupling spectroscopy signals as shown in Fig. 4a. After application of denoising and HPC models to predict possible local periods, we further apply a deep learning-based regression model to restrict the candidate hyperfine parameters for a subsequent fine-tuning process. Since the period information from HPC model only provides one functional relation between $A$ and $B$ given as Eqn (4), the purpose of the regression model is to find specific ($A$, $B$) values that best explain the shape of the coherence dips as a function of $\tau$. We set a search region for the value $B$ ranging from 10 to 70 kHz and find the best fitted ($A$, $B$) pairs repetitively for all predicted periods. Since these values are obtained by fitting coherence dips stemming from individual nuclear spins, we use the deep learning-based fit results as an initial guess and iteratively tune ($A$, $B$) pairs again in the final step to automatically search a collective list of best fitted ($A$, $B$) pairs. We describe a pseudo-code of the fine-tuning method in the Supplementary Note 2.

We demonstrate the performance of the developed procedures with two experimental datasets with $N = 32$ and $N = 256$. This new set of data was collected following the methods described in the Ref. 22 and in Supplementary Fig. 4. Figure 4b shows the comparisons of the experimental data to the reproduced CPMG signal using predicted hyperfine parameters by our deep learning protocol. Panels in Fig. 4c show example cases of predicted spins along with corresponding raw experimental data. The first panel highlights the case where the model can capture the nuclear spin signal and determine ($A$, $B$)/ $2\pi$ = (-213.19(5), 4.2(9)) (kHz) even with overlapping signals stemming from other spins. The second and third panel show that the model can accurately distinguish spins with similar periods and automatic fine tuning successfully identifies individual ($A$, $B$) pairs matching the experiments.

Our analysis returns a total of 48 nuclear spins that together accurately describe the data. However, several of these spins yield near-identical hyperfine parameters. It cannot be excluded that those signals originate from a single spin with a broadened signal due to dephasing and nuclear-nuclear spin interactions, which are not included in the model used here (see Supplementary Note 1 and Supplementary Fig.5-7 for details). We anticipate that improved selectivity in this regime is possible by using other pulse sequences, for example non-equally spaced dynamical decoupling sequence[36,58] or by taking nuclear-nuclear interactions into account. Here, we chose to count groups of spins with nearly identical parameters as a single spin. In that way, we identify 31 nuclear spins. We summarize the full list of detected nuclear spins and the confidence levels in Supplementary Table 2.

**Discussion**

We compare our result with those obtained by other methods on the same sample. A manual analysis on a similar data set, taken with the same measurement procedure, identified 7 spins with parameters that match closely to 7 of the 31 spins identified here[22]. The large improvement in number of identified spins from equivalent experimental data highlights the advantage of our deep learning approach. Additionally, we compare the results to a recent multi-dimensional spectroscopy characterization[19], a more demanding experimental technique that accesses nuclear-nuclear interactions. For 23 of the 31 spins a good match is observed (Supplementary Table 2). The other 8 spins were not previously identified and are in a spectral range that was not accessed in previous experiments. We corroborate the identification of these spins through additional experiments with a different number of decoupling pulses $N$ = 96 and $N$ = 128 (see Supplementary Fig. 8). These results show the capability of our deep learning protocol to automatically and accurately identify nuclear spins in complex spin systems and characterize the coupling parameters from dynamical decoupling spectroscopy.

We estimate a total computational time of ~3 hours from generating training datasets and training the HPC models to complete the analysis on one set of experimental CPMG data (see details in Supplementary Fig. 2). Once trained, each HPC model can identify the most probable local periods of nuclear spins from the experimental CPMG data almost instantaneously (< 1 second) and obtain the final fitted hyperfine parameters within ~50 seconds per spin (detailed specifications of the computational power used is given in Supplementary Note 1). This fast data analysis highlights the potential of deep learning approaches to efficiently scale up the sensing and characterization of large spin systems.

We find that examining dynamical decoupling spectroscopy signals for various numbers of pulses $N$ is important for the following reasons. First, large $N$ makes spins with small $B$ values visible and this is in general reflected in an increased number of detected spins as shown, for example, in the fifth panel of Fig. 4c. Second, we find that some spins near the Larmor frequency with relatively high $B$ values (> 10kHz) are detectable only in $N = 32$ since for larger $N$ too many spin signals are overlapped as illustrated in the sixth panel of Fig. 4c. The current protocol does not take nuclear – nuclear spin interactions into account. Therefore, our model fails to detect some of the interacting nuclear spins for $N = 256$, as for large $N$ and long total evolution times, nuclear-nuclear spin interactions are non-negligible and lead to a deviating period in the signal (see Supplementary Fig.9 and note that part of the nuclear-nuclear interactions are known from ref. 19). For $N = 32$, the data approximately follows a simple electron-nuclear interaction model and nuclear-nuclear interactions can be neglected. In that case, our protocol successfully detects these spins as shown in the fourth panel of Fig. 4c. We envision future improvement of the deep learning protocol by taking into account the nuclear – nuclear interactions and building a single unified model which covers all ranges of hyperfine parameters and various $N$ pulse sequences. At this current stage, discrepancies between experimental data for different $N$, for example between $N = 32$ and $N = 256$, can be used as a signature of nuclear – nuclear interaction.

In conclusion, we have proposed and demonstrated a deep learning approach to automatically detect and characterize individual nuclear spins based on dynamical decoupling spectroscopy with a single electron spin sensor. We have tested the method on a single NV center in diamond and have identified 31 individual $^{13}$C nuclear spins with a wide range of hyperfine parameters. Our method is able to distinguish spins with strong couplings to the NV center which are difficult to handle for conventional peak detection algorithms[36,59]. Our methodology retains the general benefits of deep learning models; it is easy to modify the training sequence or neural network structures for other types of experimental data such as spectroscopy data of other defect centers, including in diamond. Additionally, these results highlight the capacity of deep learning algorithms to efficiently analyze the complex nonlinear signatures in nano-scale and single-spin magnetic resonance and its robustness against realistic distortions, such as experimental decoherence and noise. Therefore, our results address one of the main challenges for quantum sensing experiments on more complex spin structures and for larger quantum registers and quantum networks based on spin qubits.

**Methods**

**Sample and setup**

Our experiments are performed on a single, naturally occurring, NV centre in a high-purity chemical-vapor-deposition homoexpitaxially grown diamond (type IIa) with a natural abundance of $^{13}$C (1.1 %) and a $\langle 111 \rangle$ crystal orientation. To improve the photon-collection efficiency, we fabricate a solid immersion lens on top of the NV center and we use an aluminum-oxide anti-reflection coating layer (grown by atomic-layer-deposition)[60]. We use on-chip lithographically-defined striplines to apply microwave fields for fast driving of the electron spin transitions (Rabi frequency ≈ 14 MHz).

We apply a static magnetic field, $B_z$ ≈ 403 G, along the NV-axis using a permanent room-temperature neodymium magnet. We stabilize the magnet field strength to < 3 mG [19] and the magnet is aligned to the NV-axis with uncertainty of 0.07° using thermal echo sequences (see ref. 19 for details of the alignment procedure).

Our experiments are performed at a temperature of 3.7 K in a commercial closed cycle cryostat

(Montana Cryostation). This enables us to readout the NV electron spin state in a single shot with high fidelity (94.5%), through spin-selective resonant excitation[60] (see detailed pulse sequence in Supplementary Fig. 4). The electron spin relaxation time is $T_1 > 1$ hour[22], the natural dephasing time is $T_2^* = 4.9(2)$ μs, the spin-echo coherence time is $T_2 = 1.182(5)$ ms, and the multipulse dynamical decoupling coherence time is $T_2^{DD} > 1$ s, for an optimized inter-pulse delay $2\tau$ [22].

**Training datasets and hyperparameters of the HPC and regression model**

For preparing training datasets of HPC and regression models, the spins of target period are chosen from ($A$, $B$) candidates separately grouped by target periods (see Supplementary Fig. 1 and Supplementary Note 1). The rest of the spin candidates are randomly selected from a previously published list calculated by DFT[60] and from a range -50 kHz < A < 50 kHz and B < 80 kHz for N32, and 2 kHz < B < 15 kHz for N256 signal (see Supplementary Fig. 2 for more details). The reason to choose this range is that the spins with larger values are already calculated in the DFT list. A range of target periods covered by one model is 250 Hz in N32 and 150 Hz in N256 with an evaluation step size of 50 Hz. (see more detailed parameter usage and performance results in Supplementary Note 1)

Although a two-dimensional convolution layer is generally employed for image recognition, to boost computational speed while retaining the accuracy and validation loss, we use Dense layers for HPC and regression models. The LeakyRelu activation shows slightly better performance for the convergence to the lower validation loss than using ReLU activation. Batch Normalization layer[54] with epsilon=1e-05, momentum=0.1 (default values in Pytorch 1.3.1) shows faster convergence to the minimum loss than Dropout regularization. For the last layer, Sigmoid layer generally converges to the higher accuracy than the Softmax layer for our datasets.

Based on the prediction results by the HPC models, the regression model is built and trained. The input data are generated from randomly selected ($A$, $B$) values of predicted periods by HPC models and the output data, in this case, are those spin values themselves (Fig. 4a). Then, the regression model is trained to predict (*A*, *B*) values of the input data. The evaluated (*A*, *B*) lists for the experimental data are used as initial guess for auto fine-tuning model.

**Training datasets and hyperparameters of the denoising model**

Unlike the HPC and regression models, training datasets for denoising models are generated in one dimensional data. The spins are selected in the same way as generating datasets for HPC models. We introduce the auto-encoder structure (Fig. 3a) which is an established structure to encode the distribution of the input data and generate the targeted data. For both encoder and decoder parts, 1D CNN layer and 1D transposed CNN layer are employed rather than RNN layers such as LSTM, GRU layers, which show lower validation errors and faster convergence to the minimum loss. All the kernel size for both CNN layers is 4. In the encoder part, Maxpooling1D layer with kernel size of 2 is used after every single 1D CNN layer. A Batch Normalization layer with default parameters is used between all CNN layers.

In all models, AdaBound[62] is employed for the optimizer and the initial learning rate is 0.00015 decayed at each epoch with customized rate (0.5 ~ 0.25). For loss functions, BinaryCrossEntropy loss is used for HPC model and MeanSquareError loss is used for regression and denoising models.

**Fine-tuning algorithm**

For the auto fine-tuning, the objective functions are defined as,

$$Loss = \sum_i^M \sum_{t \in T_i \backslash T_{bath}} [g(t) - p(t)]^2 \tag{6}$$

where g(t) is the generated CPMG signal with all predicted (*A*, *B*) values, p(t) is the experimental data, $T_i$ is a set of data points near dips of $i^{th}$ spin in the generated data and $T_{bath}$ is a set of data points where spin baths are formed. For number of predicted spins M, the optimization process runs over the 2M-dimensional space (2M variables; M for *A* and M for *B* independently).

Since the loss function is not convex, instead of adopting algorithms such as a gradient descent, particle swarm optimization (PSO)[63] is employed for the optimization process. The number of needed particles increases exponentially as search dimensions increase, yielding a large computational cost. To reduce the cost, the optimization process is conducted sequentially only on one pair of (*A*, *B*) at once and iterated for every single spin until the total loss is converged because, as in Eqn. (6), the loss function can be split into each term of $i^{th}$ spin candidate. (see pseudo code in Supplementary Note 2)

**Usage of trained models and management of total computational time**

The denoising model can be reused for other samples if the number of unit CPMG pulse sequence (*N*) and measurement time resolution are kept the same. The classifier model can be reused if the external magnetic field, *N*, measurement time resolution, and total measurement time length remain the same.

All HPC and denoising models can be trained separately and generating datasets can also be processed independently. Therefore, for example to reduce total computational time to one-third, three computers can be used independently by dividing the training regions of all $TP_k$ into three regions. More detailed parameter usage and whole procedures for implementations of each model and processing experimental data are described in Supplementary Fig. 2 and Supplementary Note 1.

**Data Availability**

Complete training datasets analyzed and utilized in this work are available upon request. Partial datasets are available at https://github.com/kyunghoon-jung/CPMG_Analysis.

**Code Availability**

All the codes required for training the proposed models and executing the fine-tuning model along with documentation are available at https://github.com/kyunghoon-jung/CPMG_Analysis.

**Acknowledgement**

This work was supported by the National Research Foundation of Korea (NRF) Grant funded by the Korean Government (MSIT) (No.2018R1A2A3075438, No.2019M3E4A1080144, No.2019M3E4A1080145) and the Creative-Pioneering Researchers Program through Seoul National University (SNU). This work was supported by the Netherlands Organisation for Scientific Research (NWO/OCW) through a Vidi grant, and as part of the Quantum Software Consortium programme (Project No. 024.003.037/3368) and the NWA-ORC program (Project No. NWA.1160.18.208). This project has received funding from the European Research Council (ERC) under the European Union's Horizon 2020 research and innovation programme (grant agreement No. 852410). This project (QIA) has received funding from the European Union's Horizon 2020 research and innovation programme under grant agreement No 820445. We thank R. Zia, M. Scheer, J. Randall and G. L. van de Stolpe for useful discussions.

**Author contributions**

K. J. designed datasets, deep learning models, and implemented the main computational procedures. M. H. A. and T. H. T. provided experimental data and theoretical backgrounds. J. Y., H. O, and H. A. provided theoretical and technical support. G. K. designed and conducted the implementation of the fine-tuning model. D. K. and T. H. T. conceived and supervised the project. All authors contributed to the preparation of the manuscript.

**Competing interest**

The authors declare no competing interests.

# References

1	Childress, L. *et al.* Coherent dynamics of coupled electron and nuclear spin qubits in diamond. *Science* **314**, 281-285 (2006).

2	Kolkowitz, S., Unterreithmeier, Q. P., Bennett, S. D. & Lukin, M. D. Sensing Distant Nuclear Spins with a Single Electron Spin. *Phys. Rev. Lett.* **109**, 137601 (2012).

3	Taminiau, T. H. *et al.* Detection and Control of Individual Nuclear Spins Using a Weakly Coupled Electron Spin. *Phys. Rev. Lett.* **109**, 137602 (2012).

4	Zhao, N. *et al.* Sensing single remote nuclear spins. *Nat. Nanotechnol.* **7**, 657-662 (2012).

5	Shi, F. Z. *et al.* Sensing and atomic-scale structure analysis of single nuclear-spin clusters in diamond. *Nat. Phys.* **10**, 21-25 (2014).

6	Taminiau, T. H., Cramer, J., van der Sar, T., Dobrovitski, V. V. & Hanson, R. Universal control and error correction in multi-qubit spin registers in diamond. *Nat. Nanotechnol.* **9**, 171-176 (2014).

7	Gao, W. B., Imamoglu, A., Bernien, H. & Hanson, R. Coherent manipulation, measurement and entanglement of individual solid-state spins using optical fields. *Nat. Photon.* **9**, 363-373 (2015).

8	Liu, G. Q. *et al.* Single-Shot Readout of a Nuclear Spin Weakly Coupled to a Nitrogen-Vacancy Center at Room Temperature. *Phys. Rev. Lett.* **118**, 150504 (2017).

9	Bernardi, E., Nelz, R., Sonusen, S. & Neu, E. Nanoscale Sensing Using Point Defects in Single-Crystal Diamond: Recent Progress on Nitrogen Vacancy Center-Based Sensors. *Crystals* **7**, 124 (2017).

10	Zopes, J. *et al.* Three-dimensional localization spectroscopy of individual nuclear spins with sub-Angstrom resolution. *Nat. Commun.* **9**, 4678 (2018).

11	Pfender, M. *et al.* High-resolution spectroscopy of single nuclear spins via sequential weak measurements. *Nat. Commun.* **10**, 594 (2019).

12	Nagy, R. *et al.* High-fidelity spin and optical control of single silicon-vacancy centres in silicon carbide. *Nat. Commun.* **10**, 1954 (2019).

13	Metsch, M. H. *et al.* Initialization and Readout of Nuclear Spins via a Negatively Charged Silicon-Vacancy Center in Diamond. *Phys. Rev. Lett.* **122**, 190503 (2019).

14	Cujia, K. S., Boss, J. M., Herb, K., Zopes, J. & Degen, C. L. Tracking the precession of single nuclear spins by weak measurements. *Nature* **571**, 230–233 (2019).

15	Nguyen, C. T. *et al.* Quantum Network Nodes Based on Diamond Qubits with an Efficient Nanophotonic Interface. *Phys. Rev. Lett.* **123**, 183602 (2019).

16	Hensen, B. *et al.* A silicon quantum-dot-coupled nuclear spin qubit. *Nat. Nanotechnol.* **15**, 13-17 (2020).

17	Muller, C. *et al.* Nuclear magnetic resonance spectroscopy with single spin sensitivity. *Nat.*

*Commun.* **5**, 4703 (2014).

18   Zopes, J., Herb, K., Cujia, K. S. & Degen, C. L. Three-Dimensional Nuclear Spin Positioning Using Coherent Radio-Frequency Control. *Phys. Rev. Lett.* **121**, 170801 (2018).

19   Abobeih, M. H. *et al.* Atomic-scale imaging of a 27-nuclear-spin cluster using a quantum sensor. *Nature* **576**, 411-415 (2019).

20   Yang, Z. *et al.* Structural Analysis of Nuclear Spin Clusters via 2D Nanoscale Nuclear Magnetic Resonance Spectroscopy. *Adv. Quantum Technol.* **3**, 1900136 (2020).

21   Yao, N. Y. *et al.* Scalable architecture for a room temperature solid-state quantum information processor. *Nat. Commun.* **3**, 800 (2012).

22   Abobeih, M. H. *et al.* One-second coherence for a single electron spin coupled to a multi-qubit nuclear-spin environment. *Nat. Commun.* **9**, 2552 (2018).

23   Humphreys, P. C. *et al.* Deterministic delivery of remote entanglement on a quantum network. *Nature* **558**, 268-273 (2018).

24   Bradley, C. E. *et al.* A Ten-Qubit Solid-State Spin Register with Quantum Memory up to One Minute. *Phys. Rev. X* **9**, 031045 (2019).

25   Hou, P. Y. *et al.* Experimental Hamiltonian Learning of an 11-Qubit Solid-State Quantum Spin Register. *Chin. Phys. Lett.* **36**, 100303 (2019).

26   van der Sar, T. *et al.* Decoherence-protected quantum gates for a hybrid solid-state spin register. *Nature* **484**, 82-86 (2012).

27   Waldherr, G. *et al.* Quantum error correction in a solid-state hybrid spin register. *Nature* **506**, 204-207 (2014).

28   Cramer, J. *et al.* Repeated quantum error correction on a continuously encoded qubit by real-time feedback. *Nat. Commun.* **7**, 11526 (2016).

29   Unden, T. *et al.* Quantum Metrology Enhanced by Repetitive Quantum Error Correction. *Phys. Rev. Lett.* **116**, 230502 (2016).

30   van Dam, S. B., Cramer, J., Taminiau, T. H. & Hanson, R. Multipartite Entanglement Generation and Contextuality Tests Using Nondestructive Three-Qubit Parity Measurements. *Phys. Rev. Lett.* **123**, 050401 (2019).

31   Unden, T. K., Louzon, D., Zwolak, M., Zurek, W. H. & Jelezko, F. Revealing the Emergence of Classicality Using Nitrogen-Vacancy Centers. *Phys. Rev. Lett.* **123**, 140402 (2019).

32   Kalb, N. *et al.* Entanglement distillation between solid-state quantum network nodes. *Science* **356**, 928-932 (2017).

33   Rozpędek, F. *et al.* Optimizing practical entanglement distillation. *Phys. Rev. A* **97**, 062333 (2018).

34   Zhao, N., Wrachtrup, J. & Liu, R.-B. Dynamical decoupling design for identifying weakly coupled nuclear spins in a bath. *Phys. Rev. A* **90**, 032319 (2014).

35  Hürlimann, M. D., Utsuzawa, S. & Hou, C.-Y. Spin Dynamics of the Carr-Purcell-Meiboom-Gill Sequence in Time-Dependent Magnetic Fields. *Phys. Rev. Appl.* **12**, 044061 (2019).

36  Casanova, J., Wang, Z. Y., Haase, J. F. & Plenio, M. B. Robust dynamical decoupling sequences for individual-nuclear-spin addressing. *Phys. Rev. A* **92**, 042304 (2015).

37  Jelezko, F. in *Quantum Information Processing with Diamond* (eds Steven Prawer & Igor Aharonovich) xxi-xxii (Woodhead Publishing, 2014).

38  Bock, K. & Pedersen, C. in *Advances in Carbohydrate Chemistry and Biochemistry* Vol. 41 (eds R. Stuart Tipson & Derek Horton) 27-66 (Academic Press, 1983).

39  Tognarelli, J. M. *et al.* Magnetic Resonance Spectroscopy: Principles and Techniques: Lessons for Clinicians. *J. Clin. Exp. Hepatol.* **5**, 320-328 (2015).

40  Wurz, J. M., Kazemi, S., Schmidt, E., Bagaria, A. & Guntert, P. NMR-based automated protein structure determination. *Arch. Biochem. Biophys.* **628**, 24-32 (2017).

41  Guntert, P. Automated structure determination from NMR spectra. *Eur. Biophys. J. Biophy.* **38**, 129-143 (2009).

42  Abbas, A., Kong, X. B., Liu, Z., Jing, B. Y. & Gao, X. Automatic Peak Selection by a Benjamini-Hochberg-Based Algorithm. *Plos One* **8**, e53112 (2013).

43  Polanski, A., Marczyk, M., Pietrowska, M., Widlak, P. & Polanska, J. Signal Partitioning Algorithm for Highly Efficient Gaussian Mixture Modeling in Mass Spectrometry. *Plos One* **10**, e0134256 (2015).

44  Nerli, S., McShan, A. C. & Sgourakis, N. G. Chemical shift-based methods in NMR structure determination. *Prog. Nucl. Mag. Res. Sp.* **106**, 1-25 (2018).

45  Klukowski, P. *et al.* NV center based nano-NMR enhanced by deep learning. *Bioinformatics* **34**, 2590-2597 (2018).

46  Aharon, N. *et al.* NMRNet: a deep learning approach to automated peak picking of protein NMR spectra. *Sci. Rep.* **9**, 17802 (2019).

47  Vilar, D., Castro, M. J. & Sanchis, E. Multi-label text classification using multinomial models. *Lect. Notes Artif. Int.* **3230**, 220-230 (2004).

48  Godbole, S. & Sarawagi, S. Discriminative methods for multi-labeled classification. *Lect. Notes Artif. Int.* **3056**, 22-30 (2004).

49  Boutell, M. R., Luo, J. B., Shen, X. P. & Brown, C. M. Learning multi-label scene classification. *Pattern Recognit.* **37**, 1757-1771 (2004).

50  Ou, G. B. & Murphey, Y. L. Multi-class pattern classification using neural networks. *Pattern Recognit.* **40**, 4-18 (2007).

51  Liu, S. M. & Chen, J. H. A multi-label classification based approach for sentiment classification. *Expert Syst. Appl.* **42**, 1083-1093 (2015).

52  Lecun, Y., Bottou, L., Bengio, Y. & Haffner, P. Gradient-based learning applied to document

recognition. *P. IEEE* **86**, 2278-2324 (1998).

53  He, K., Zhang, X., Ren, S. & Sun, J. Deep Residual Learning for Image Recognition. Preprint at https://arxiv.org/abs/1512.03385. (2015).

54  Ioffe, S. & Szegedy, C. Batch Normalization: Accelerating Deep Network Training by Reducing Internal Covariate Shift. Preprint at https://arxiv.org/abs/1502.03167. (2015).

55  Vincent, P., Larochelle, H., Lajoie, I., Bengio, Y. & Manzagol, P. A. Stacked Denoising Autoencoders: Learning Useful Representations in a Deep Network with a Local Denoising Criterion. *J. Mach. Learn. Res.* **11**, 3371-3408 (2010).

56  Baldi, P. in *Proceedings of ICML Workshop on Unsupervised and Transfer Learning* Vol. 27 (eds Guyon Isabelle *et al.*) 37-49 (PMLR, Proceedings of Machine Learning Research, 2012).

57  Kiranyaz, S. *et al.* 1D Convolutional Neural Networks and Applications – A Survey. Preprint at https://arxiv.org/abs/1905.03554. (2019).

58  Wang, Z. H., de Lange, G., Riste, D., Hanson, R. & Dobrovitski, V. V. Comparison of dynamical decoupling protocols for a nitrogen-vacancy center in diamond. *Phys. Rev. B* **85**, 155204 (2012).

59  Oh, H. S. *et al.* Algorithmic decomposition for efficient multiple nuclear spin detection in diamond. Preprint at https://arxiv.org/abs/2003.00178. (2020).

60  Robledo, L. *et al.* High-fidelity projective read-out of a solid-state spin quantum register. *Nature* **477**, 574-578 (2011).

61  Nizovtsev, A. P. *et al.* Non-flipping C-13 spins near an NV center in diamond: hyperfine and spatial characteristics by density functional theory simulation of the C-510[NV]H-252 cluster. *New. J. Phys.* **20**, 023022 (2018).

62  Luo, L., Xiong, Y., Liu, Y. & Sun, X. Adaptive Gradient Methods with Dynamic Bound of Learning Rate. Preprint at https://arxiv.org/abs/1902.09843. (2019).

63  Kennedy, J. & Eberhart, R. in *Proceedings of ICNN'95 - International Conference on Neural Networks*. 1942-1948 (1995).

# Supplementary Information

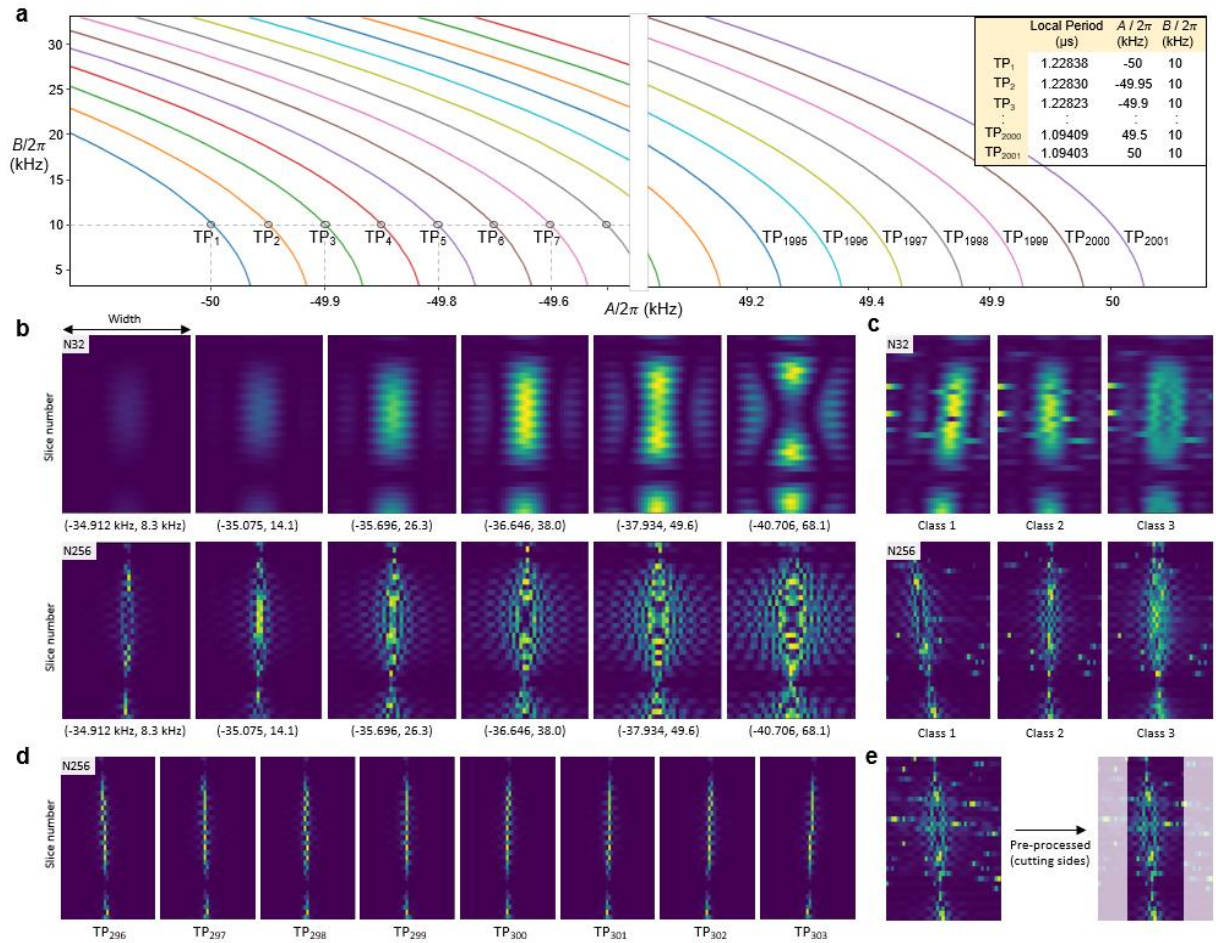

**Supplementary Figure 1. Range of hyperfine parameters (*A*, *B*) used for proposed models. a.** Contour lines of local periods ($TP_k$) in the (*A*, *B*) - plane. All (*A*, *B*) pairs on one contour line have the same local period but show different fringe patterns. For each hyperfine parameter classifier (HPC) model, the training dataset is generated by using (*A*, *B*) pairs on each $TP_k$ curve. For example, the inset table displays *TP*s calculated with (*A*, *B*) values at black circled points using the Eqn. (4) in the main text. **b.** Variation of central dips and oscillatory fringe patterns change depending on (*A*, *B*) values in the group $TP_{300}$ for pulse repetition *N* = 32 (top row) and *N* = 256 (bottom row). From left to right panels, the local period of each panel remains the same while *B* value increases, showing more complicated characteristics. **c.** Example input images for no, single, and double target spins in HPC model for (*A*, *B*) values in group $TP_{300}$. **d.** Panels from left to right show the slope for a single nuclear spin (*A*: -34.85 kHz, *B*: 6 kHz) as a function of varying target periods in step of 50 Hz in *A* (from $TP_{296}$ to $TP_{303}$). The width of all images in b, c and d panels is 140 ns and the total number of slices is 33. The colour scale bar in all the images ranges from 0 to 1. **e.** Data pre-processing by cutting both sides of an image for HPC models to improve the classification performance and to reduce computational cost.

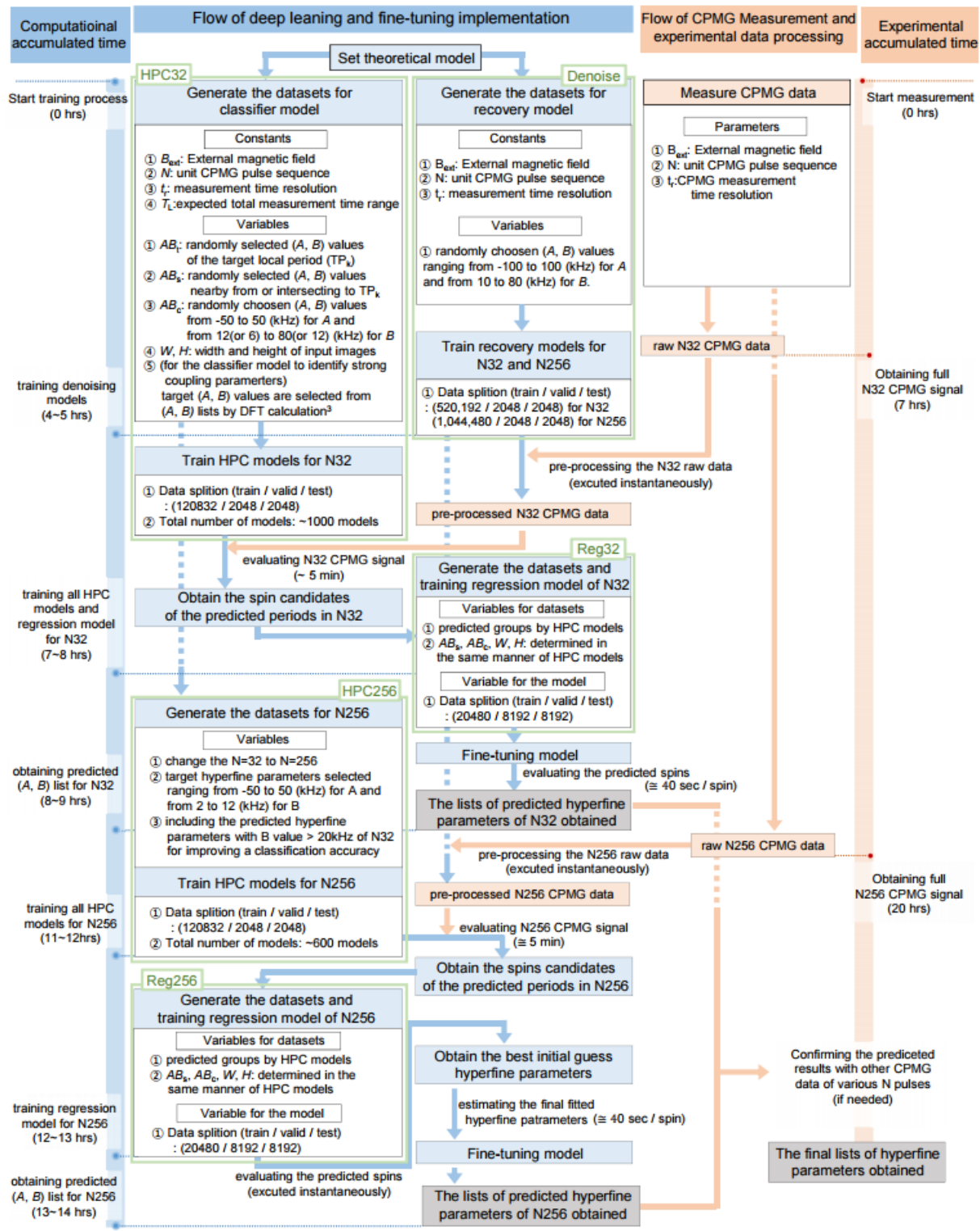

**Supplementary Figure 2. Flow chart of the deep learning protocol and experimental CPMG measurement.** The total procedure has two main streams; one is for preparing datasets and training proposed models (blue stream) and the other is for measuring experimental CPMG signals (pink stream). The timeline displayed in left and right sides are the total accumulated time for each flow. Each essential step is marked by light-green rectangular boxes denoted as Denoise, HPC32(256), and Reg32(256). Details are described in Supplementary Note. 1.

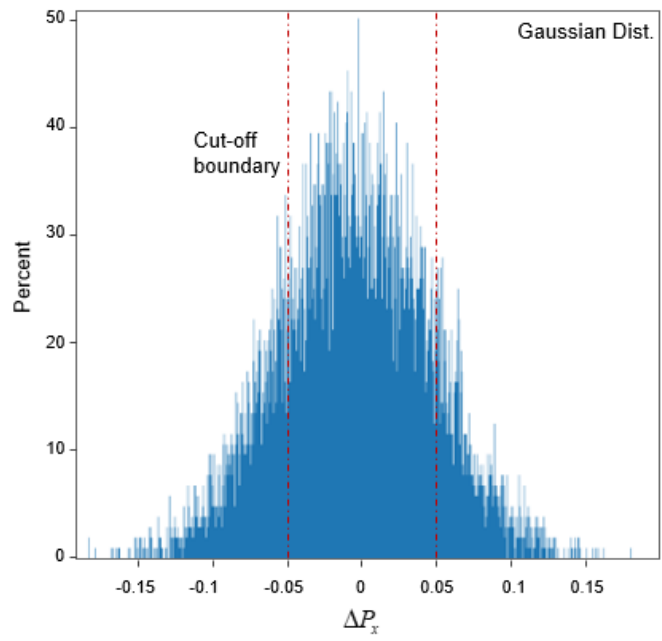

**Supplementary Figure 3. Histogram of the Gaussian distribution used for generating noisy training datasets.** This reflects the noise of the experimental data; a cut-off boundary with amplitude of 0.05 measured from the CPMG experimental data is used.

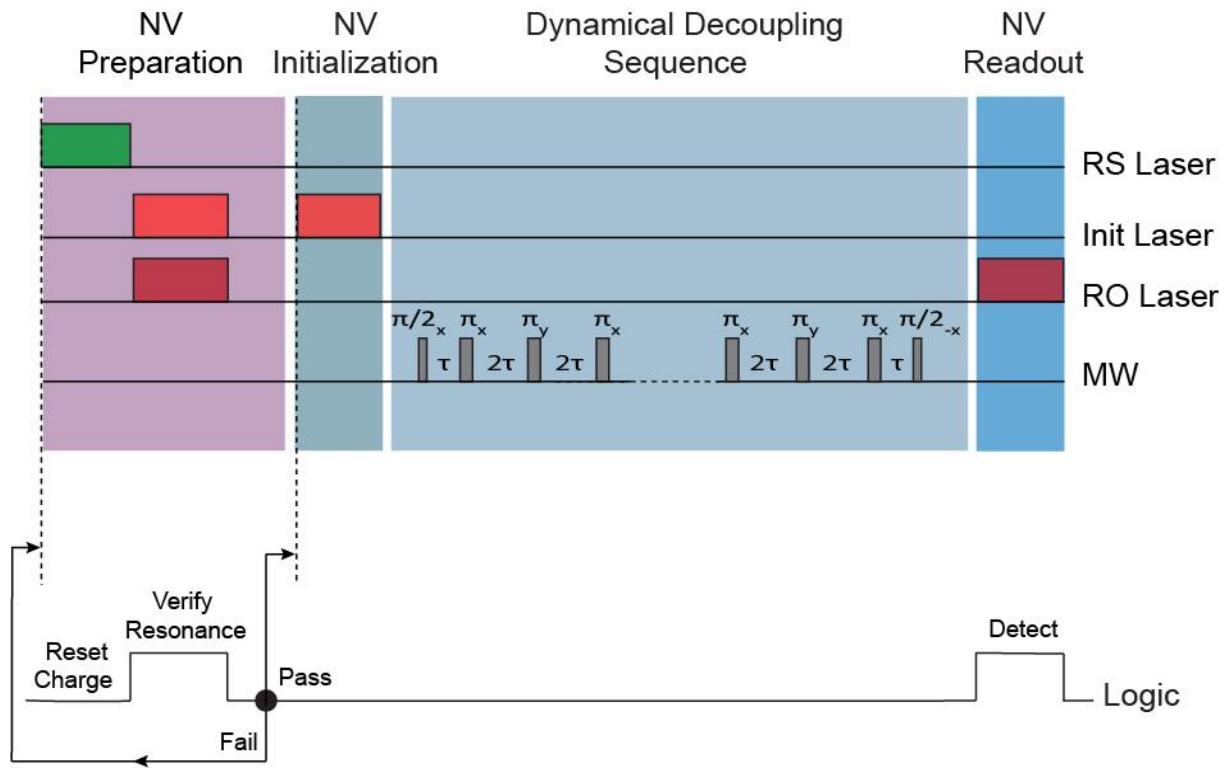

**Supplementary Figure 4. Experimental sequence.** The pulse sequence consists of four main parts: NV Preparation, NV Initialization, Dynamical Decoupling Sequence, NV Readout. **NV Preparation**: first the NV centre is prepared in the negative charge state and brought on resonance with the lasers used for the initialization (Init) and Readout (RO) steps. We simultaneously apply the Init and RO lasers for 150 μs and count the number of detected photons (wavelength ~ 637 nm, RO laser resonant with the $E_x$ transition and Init laser resonant with the E' transition) [1]. If the number of detected photons exceeds a certain threshold, the NV is in the negative charge state and on resonance with both lasers, and the sequence proceeds to the next step. If not, the charge reset laser (RS, wavelength ~ 515 nm) is applied for 1 ms and the same process is repeated until success [1]. **NV Initialization**: we initialize the NV electron spin into the $m_s = 0$ state through spin pumping into the E' transition (Init laser, 100 μs) [1]. **Dynamical Decoupling Sequence**: the electron spin is prepared into a superposition state by applying a π/2-pulse. Afterwards, a sequence of $N$ π-pulses is applied on the electron spin with the form ($\tau$ - π - π)$^N$. To reduce the effect of pulse errors, we alternate the phases of the π-pulses according to the XY-8 scheme [2]. **NV Readout**: we apply the RO laser for 10 μs and count the number of detected photons in this period. This allows us to read out the NV electron spin state in a single shot (Fidelity ~ 94.5%).

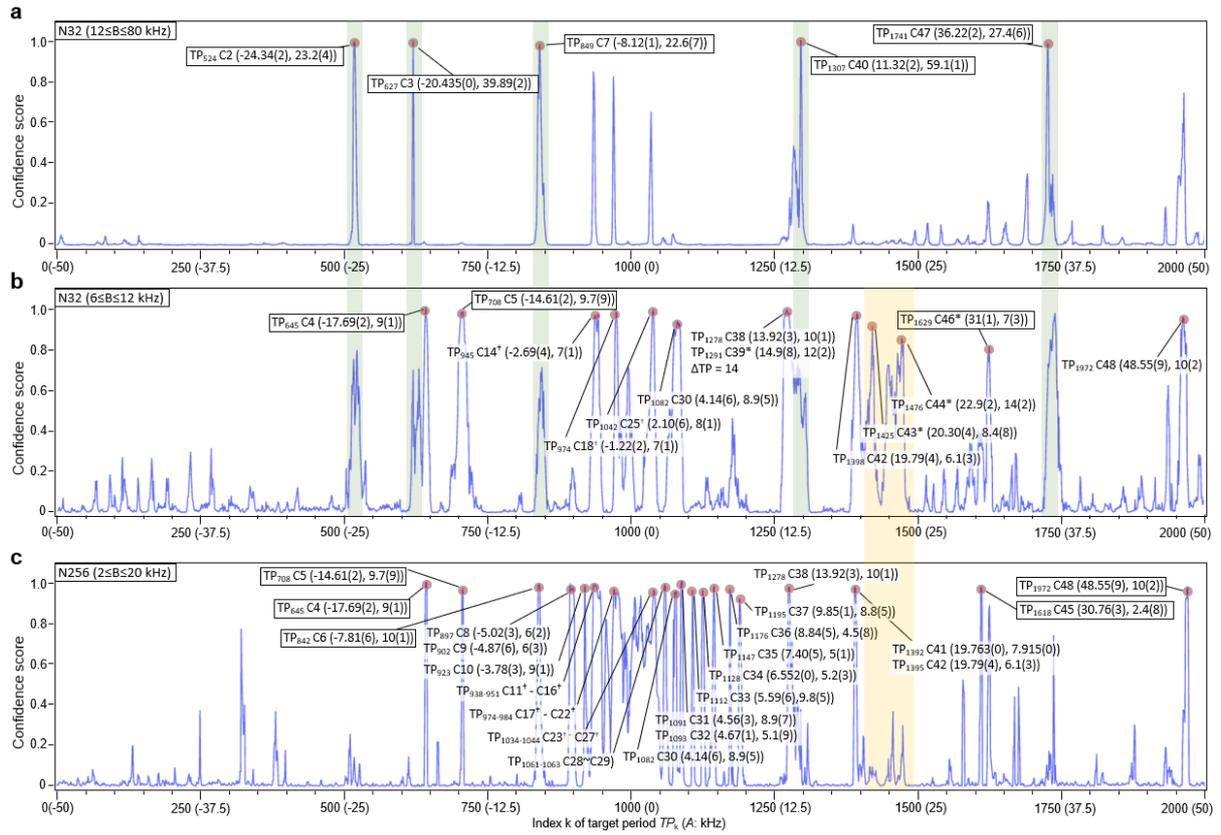

**Supplementary Figure 5. Results from HPC models. a.** Confidence score for N32 ($B \geq 12$ kHz). **b**. Confidence score for N32 ($6 \leq B \leq 12$ kHz). **c**. Confidence score for N256 ($2 \leq B \leq 20$ kHz). See Supplementary Note 1 for discussion on the validity of nuclear spin detection highlighted in green and yellow regions. All data are averaged over 10 iterations of training.

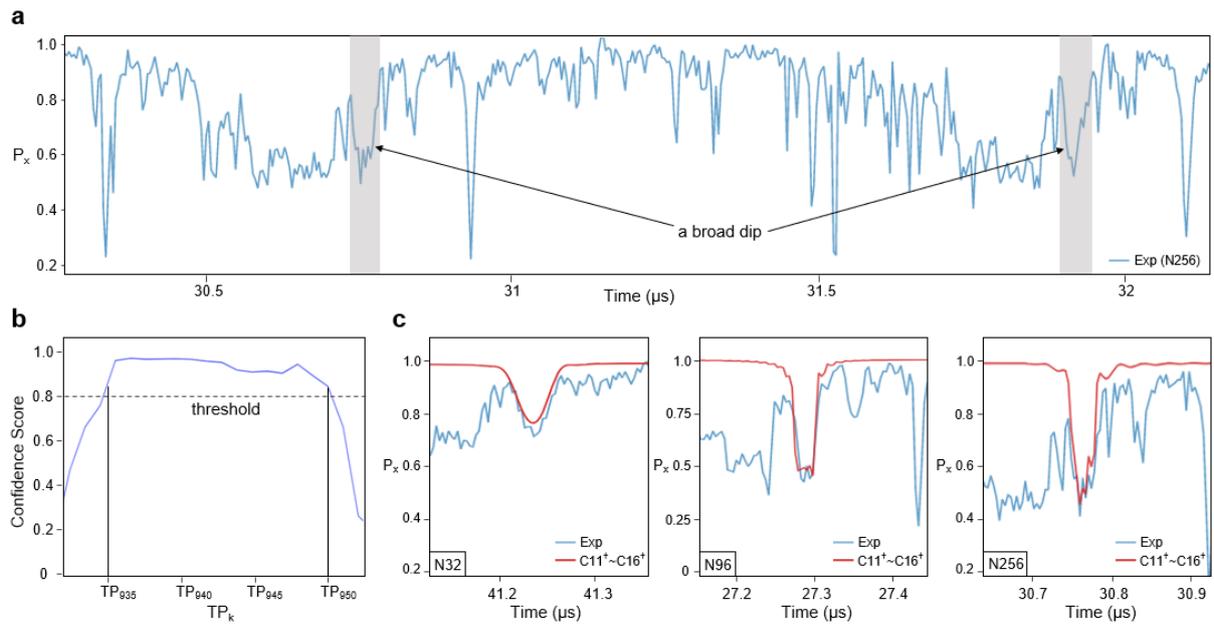

**Supplementary Figure 6. Identifying a group of spins for a single broad dip. a.** A single broad dip (coloured in grey) formed by a group of spins with similar local periods. **b.** Confidence scores for the broad dip in the panel a. We note that the confidence scores are higher than the threshold (0.8) for several target period indices. This indicates that such a broad dip might be explained by many spins with similar hyperfine parameters (see Supplementary Note 1) **c.** A comparison of a simulated signal (red line) of $C11^{\dagger} \sim C16^{\dagger}$ with the experimental CPMG signal (blue line) for different pulses N32, N96 and N256. More detailed procedures to estimate the number of spins to fit a broad dip are described in Supplementary Note 1.

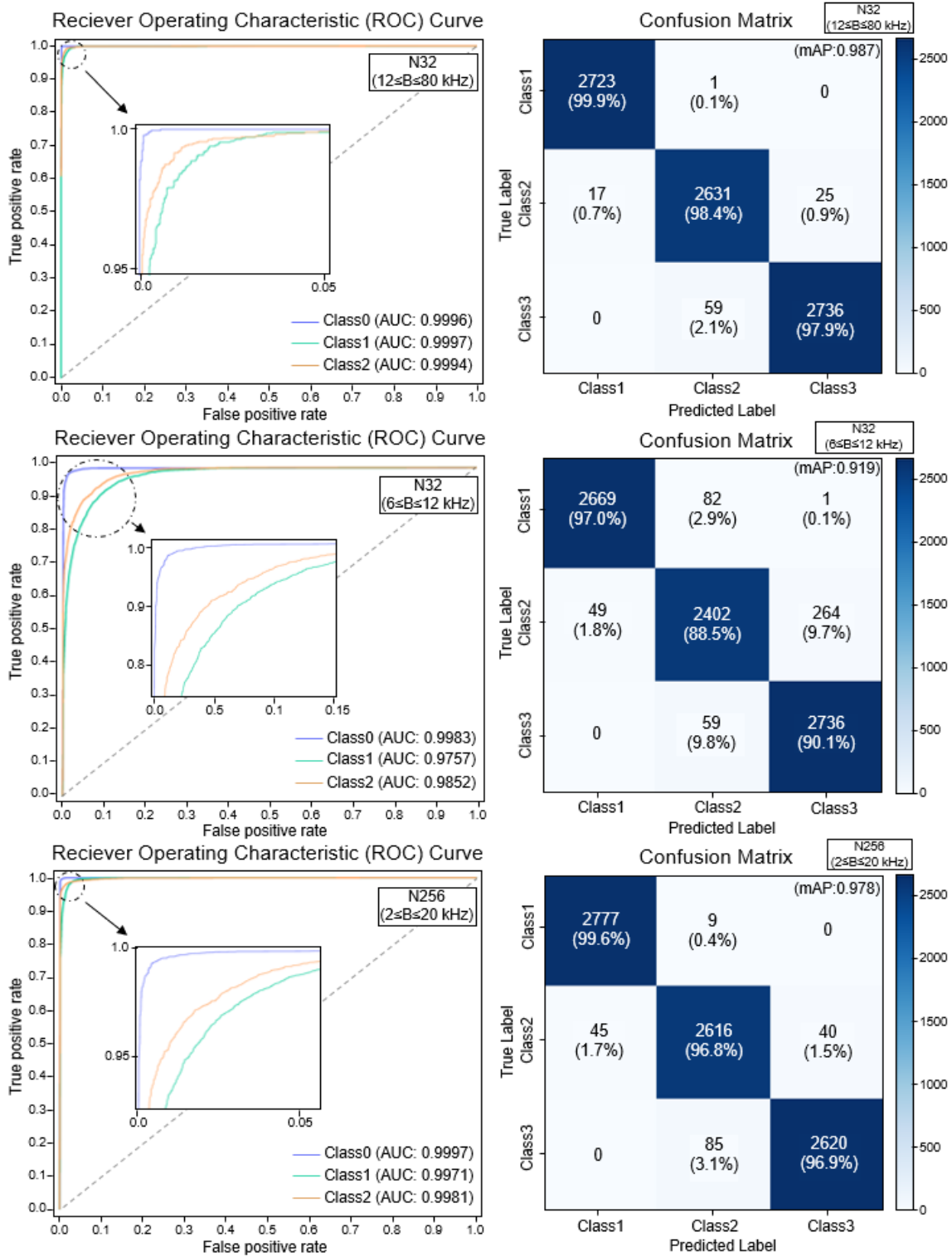

**Supplementary Figure 7. Receiver operating characteristic (ROC) curves and corresponding confusion matrices for HPC models.** All the ROC curves and confusion matrices are obtained by averaging the results of all the HPC models which cover the range (unit: kHz) of $-50 \leq A \leq 50$ (for both N32 and N256), $12 \leq B \leq 80$ (the top panel, N32), $6 \leq B \leq 12$ (the second, N32) and $2 \leq B \leq 12$ (the bottom, N256). Each ROC curve is calculated through one-vs-rest class convention. AUC indicates the area under the curve for each class. In confusion matrices, the numbers in boxes indicate the average number of samples and the number in parentheses indicates the ratio of the number of samples to the number of all the true labels for each class. mAP indicates macro-average-precision. The accuracy in the top, the second and the third cases is 98.8%, 95.3% and 97.8% respectively.

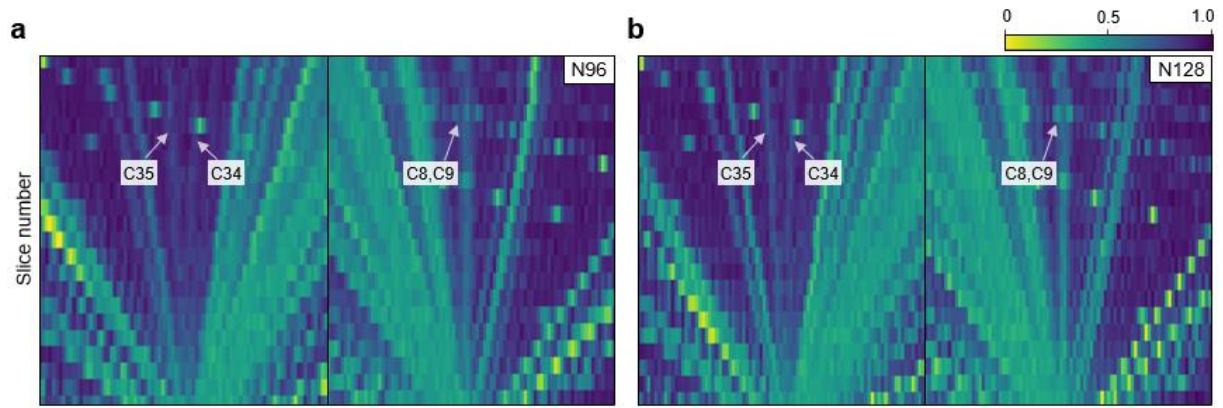

**Supplementary Figure 8. Verification of nuclear spin detection in *N*=96 and *N*=128.** In N256 signal, we detect several nuclear spins that were not observed in previous work on the same NV center sample [3] (see Supplementary Table 2) and confirm some cases using different *N* pulses. Each panel shows examples using **a.** *N* = 96 and **b.** *N* = 128 CPMG data. The width of all images is 100 ns and the total number of slices is 22.

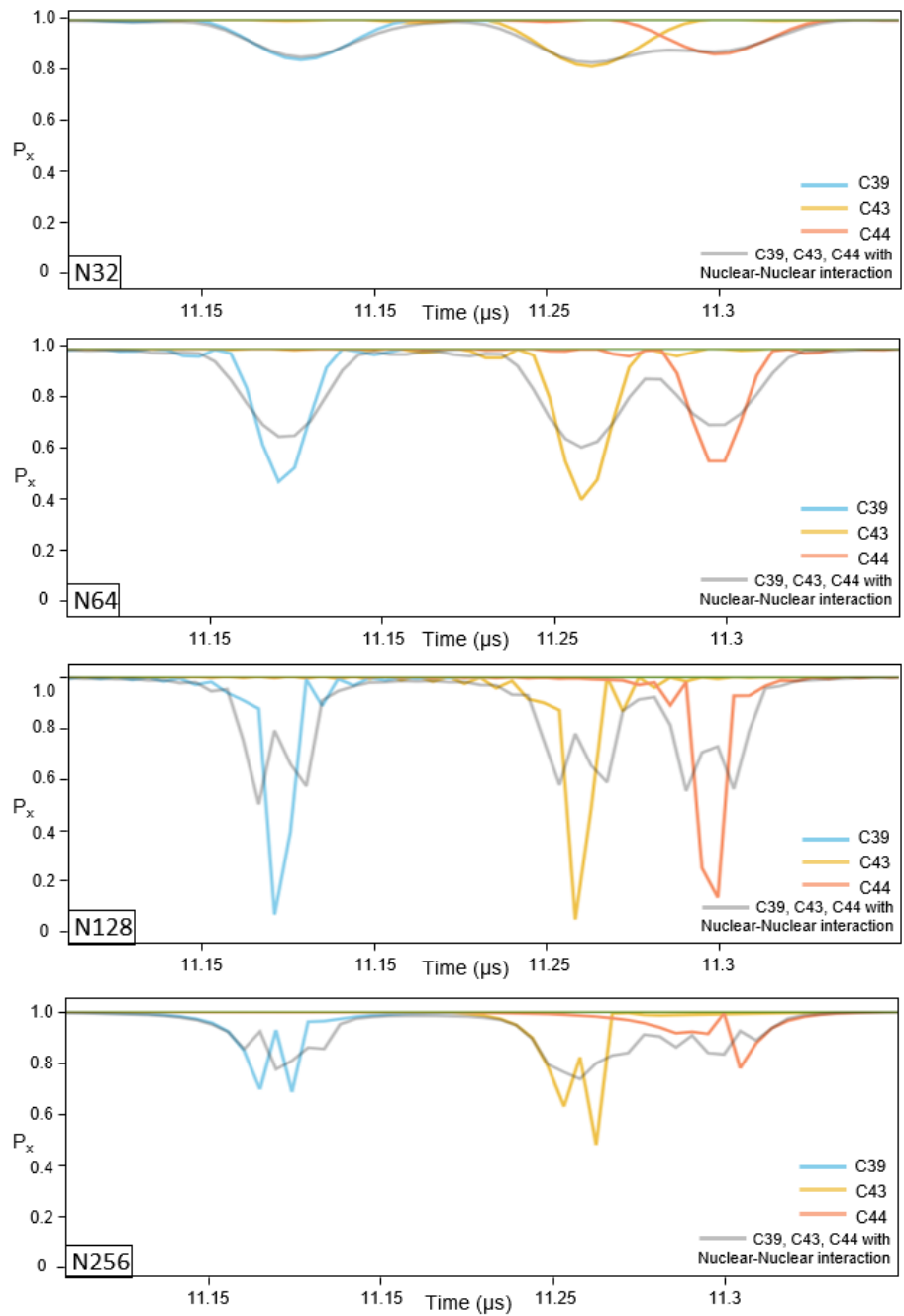

**Supplementary Figure 9. Comparison of CPMG signals with or without nuclear-nuclear interaction. Top to bottom panels :** The effect of nuclear-nuclear interaction on CPMG signal for $N = 32$ to $N = 256$ pulse sequences. Blue, yellow, and red lines are simulated by the current theoretical model and a grey line is drawn from a preliminary model including nuclear-nuclear interaction. For example, for a case of C43* and C44* in Supplementary Fig.5b and 5c, confidence scores covered with a light-yellow bar shows high confidence scores in N32 ($6 \leq B \leq 12$ kHz) but low in N256 ($2 \leq B \leq 20$ kHz). The nuclear-nuclear coupling in this case are set to 236 Hz, 62 Hz and 25 Hz. These are representative values for nuclei with large couplings, to highlight the effect. Typical nuclear-nuclear interactions in the cluster are much smaller [3].

**Supplementary Note 1. Detailed procedure of machine learning protocol.**

**Step.1 Denoising model (Annotated as Denoise in Supplementary Fig.2) – Recovering noisy signals.** To generate training datasets for a denoising model, (A, B) pairs are collected in the same way as in HPC models but not pre-processed in the image representation as in Fig.2 in the main text. We use the gaussian noise distribution with standard deviation 0.05, which reflects the maximum amplitude of experimental noise. (see Supplementary Fig. 3) The length of input data is 12 μs (3000 data points with 4 ns time resolution in our experimental setup) but the performance of the model does not depend much on the length between 1000 and 3000. The sample datasets and training implementation is described in an online tutorial section; Tutorial part 2 – Denoising model at https://github.com/kyunghoon-jung/Deep_Learning_CPMG_Analysis.

**Step.2 HPC model (HPC32, HPC256 in Supplementary Fig.2) – Identifying local periods of existing nuclear spins.** The target period ($TP_k$) shown in Supplementary Fig. 1a is calculated with the approximate equation [4],

$$TP_k = 2\pi / (\tilde{\omega}_k + \omega_L) \tag{1}$$

, where $\tilde{\omega}_k = \sqrt{(A_k + \omega_L)^2 + B_k^2}$, $\omega_L$ is the Larmor frequency and $A_k$ and $B_k$ are hyperfine parameters of the $k^{th}$ target period. We pre-calculate target periods by a step size of $A$ = 50 Hz with a fixed $B$ (=10 kHz) value as seen in the inset table in Supplementary Fig. 1a and save it as a Python dictionary file - [key ($A$ values) : value (target periods)]. The evaluation step size is 50 Hz for $A$. Supplementary Figure 1d describes how a slope of a central line changes at each step. The data manipulation and the dictionary file are presented in an online tutorial; Tutorial part 1 - Data handling and generation in the proposed representation at the same link above.

The external magnetic field ($B_z$) is 403.553 G. The measurement time resolution ($t_r$) of the CPMG signal is 4 ns. The width (W) of the input images sliced by $TP_k$ are reduced to be 80~120 ns for the range of $A \leq$ -10 kHz and $A \geq$ 10 kHz and to be 40~80 ns for -10 $\leq A \leq$ 10kHz for the optimal performance and reducing the computational cost. (see Supplementary Fig.1e) The height (H, total number of slices) is determined depending on the total measurement length of the experimental data. The total measurement length of experimental CPMG data of N32 (N256) is 6 ~ 50 μs (10 ~ 40 μs).

Spins with different periods from $TP_k$ and spins from the list in Supplementary Table 1 taken from the density functional theory (DFT) calculation result [5] are included in training datasets to make the model robust in the presence of other spins with various periods. The total number of spins for generating each input data is randomly selected between 26~32 where 90~100% of the total number of spins are selected from the range -50 $\leq A \leq$ 50 kHz, 6 $\leq B \leq$ 80 kHz for $N$ = 32 datasets and -50 $\leq A \leq$ 50 kHz, 2 $\leq B \leq$ 20 kHz for $N$ = 256 datasets while the rest 0~10% of the spins are selected from the DFT list. The reason to choose the upper bound as $B$ = 20 kHz for $N$ = 256 datasets is that the spins with $B \geq$ 20 kHz can be evaluated accurately in N32. (see Supplementary Fig.5a and the top panels in Supplementary Fig.7) When evaluating the strong coupling regime (for example $\omega_h/2\pi >$ 80 kHz), the HPC models only cover the range near the (A, B) values in the DFT list because the distribution of (A, B) values in this regime is distributed sparsely. [5] (also see Supplementary Table 1).

At first, HPC models are trained with two classes; Class 1 is generated excluding a spin with the target period and Class 2 is generated including a spin with the period. And to reduce computational cost, training datasets consist of multiple target periods so one HPC model is trained to evaluate several target periods. For example, if a model is built with datasets of $TP_{501} \sim TP_{505}$, the model is trained to evaluate five target periods ($TP_{501}$, $TP_{502}$, $TP_{503}$, $TP_{504}$ and $TP_{505}$). We train one HPC model with 5 target periods for N32 and 3 target periods for N256.

Supplementary Figure 5 shows confidence scores for N32 and N256 and detected spins circled in red. As shown in Supplementary Fig. 5a,b, when evaluating the signal of N32, we train the HPC models with two different range of $B$ values to enhance the performance for a range of small $B$ values (6 $\leq B \leq$ 12kHz). In green coloured regions in panel b, spins are not selected although some points show high confidence scores. Because target period indices of spins with large $B$ values (C2, C3, C7, C40 and C47) are overlapped in this region as seen in panel a, b.

To extract predicted periods of high confidence scores (see red circles in Supplementary Fig.5), we use the 'find_peaks' function (a signal processing function of scipy package in Python) with parameters of height=0.75(N32, $6 \leq B \leq 12$kHz) and 0.9(N32, $B \geq 12$kHz and N256), distance=4, width=2, prominence=0.5. The classification resolution is $A$ = 500 Hz for N32 ($6 \leq B \leq 12$ kHz), $A$ = 200 Hz for N32 ($B \geq 12$ kHz) and $A$ = 150 Hz for N256 of $B \leq 20$kHz; a classification resolution means the minimum difference where two different nuclear spins can be accurately distinguished. See an implementation sample in an online tutorial; Tutorial part 3, 4 – Hyperfine parameter classifier (HPC) model at the same link above.

**Step 3. Regression model (Reg32, Reg256 in Supplementary Fig. 2) – Estimating ($A$, $B$) values for predicted periods in Step.1.** Before specifying ($A$, $B$) values, we train additional HPC models to determine the number of nuclear spins for the case of a broad dip (see Supplementary Fig. 6). A broad dip in N256 shows confidence scores higher than a threshold in a range of more than four target period indices (see an example in Supplementary Fig. 6a, b). The structure of the model remains the same as in Fig.2 in the main text, but an input dataset is generated with a group of target spins selected within the predicted periods.

For example, in Supplementary Fig.6b, confidence scores of $TP_{935} \sim TP_{950}$ are higher than the threshold. In this case, an output one-hot vector has 5 dimensions where Class 1 [1,0,0,0,0] corresponds to 'No target spin', Class 2 [0,1,0,0,0] corresponds to the number of spins = 4 and Class 3 [0,0,1,0,0] (Class 4 [0,0,0,1,0], Class 5 [0,0,0,0,1]) to the number of spins = 5 (6, 7) respectively. Heuristically, one spin exists if confidence scores are shown higher than a threshold for 2 ~ 4 target periods. So, for example, if confidence scores are high for 8 target periods (or 13 target periods), then the minimum number of spins to exist is 2 (or 4). This determines the minimum number (=4 in this case because confidence scores are high for $TP_{935} \sim TP_{950}$) of spins for training datasets. The maximum number (=7 in this example) of spins is chosen depending on a classification resolution of the HPC model. The additional HPC model outputs a vector with the highest confidence score in Class 4 which is assigned to the number of spins = 6. (Output vector is [0.00012, 0.00332, 0.2312, 0.6756, 0.3121]) After obtaining the number of spins, the regression step in Fig.4 in the main text is performed.

Although the model quantitatively estimates the number of spins, the actual number of spins underlying these broader signals is difficult to be determined, as nuclear-nuclear interactions and dephasing can play a role, and are not taking into account in our model. Therefore, we choose to count a group of spins forming a broad dip (denoted as C#[†] in Supplementary Table 2) as one spin so that the groups C11[†]~C16[†], C17[†]~C22[†], C23[†]~C27[†], C31[†]~C32[†] and C41[†]~C42[†] are counted as five spins in total (each contains at least one spin). This conservative way of counting reduces the total number of spins from 48 to 31. A more detailed implementation can be found in an online tutorial; Tutorial part 5 – Regression Model at the same link above.

**Classification performance – ROC curves and confusion matrices.** For ROC curves in Supplementary Fig. 7, AUC scores in N32 ($6 \leq B \leq 12$ kHz) are lower than those in other cases because the dips of spins with $B$ ($6 \leq B \leq 12$ kHz) in N32 are less distinct than those with larger $B$ values. The lower bound of $B$ values (6 kHz in N32 and 2 kHz in N256) of HPC models are determined by the minimum amplitudes of dips distinguishable by models. The bounds can be customized for more optimal performances depending on measurement configurations. The confusion matrices show HPC models in N32 ($6 \leq B \leq 12$ kHz) have difficulty distinguishing Class 1 (one target period) and Class 2 (two target periods), which leads to a relatively low classification resolution in this range. Due to this ambiguity in the results of HPC models with this range, we count the number of spins in a conservative way for the case of continuously high confidence scores. For example, an area of high confidence scores within a range of 10 target periods is counted as one spin and for a range of 10~20 target periods the number of spins is estimated as two spins. (see the cases of C38, C39 in Supplementary Fig.5b)

**Hardware specification and computational time management** – The total accumulated computational time is shown in left side of Supplementary Fig. 2. The required time is measured by a workstation with following specifications. CPU: Intel Core i9-9920X (12 cores, 24 threads), RAM: 128GB DDR4, and GPUs: NVIDIA RTX 2080Ti (x2). In the whole procedure, the usage of multi-threads is fixed to 20 threads.

The computational time can be reduced by using multiple computers simultaneously because Step.1 and Step. 2 can be trained independently. In addition, training each HPC model and generating datasets can be implemented independently. The trained denoising model can be reused for other experimental data sets if the number of CPMG pulse sequence units ($N$) and measurement time resolution ($t_r$) are kept the same, while the trained classifier model can be reused when the $B_{\text{ext}}$, $N$, $t_{\text{r}}$, and $T_{\text{L}}$ remain the same. ($N$, $t_{\text{r}}$, $B_{\text{ext}}$, and $T_{\text{L}}$ are parameters shown in Supplementary Fig.2)

**Discussion – Effect of nuclear-nuclear interaction** In Supplementary Fig.5b, c (see light-yellow bar), it is shown that C43, C44 has high confidence scores in N32 but not in N256 because nuclear-nuclear interaction turns significant as $N$ pulse increases (see Supplementary Fig.9). Comparing the top panel (N32) and the bottom panel (N256) in Supplementary Fig.9, a grey line in N256 shows a small and smooth dip making HPC models fail to detect nuclear spins. In our current theoretical model, nuclear-nuclear interaction is not taken into account. The main difficulty in addressing this nuclear-nuclear interaction lies in computing all the possible interactions between nuclear spin candidates.

## Supplementary Note 2. Pseudo code of the fine-tuning method.

---
**Algorithm 1:** Fine Tunning Algorithm
---

**Function** `FineTunning`:
    **Input:** Regression data list $[(A_i, B_i)]$, $i = 1, 2, \cdots, n$, where HPC values are from regression.
    **Output:** Processed list.
    **for** Idx, (A, B) in enumerate(Regression_list) **do**
        P = Particles_Initialization(Regression_list, Idx, dB, N_particles)
        best_loss = loss(P[0])
        best_particle = P[0]
        **for** p in P **do**
            **if** *N_pulse == 32* **then**
                | Optimize p with Conjugate Gradient Optimizer.
            **else**
                | Optimize p with L-BFGS-B Optimizer.
            **end**
            **if** *loss(p) < best_loss* **then**
                best_loss $\leftarrow loss(p)$
                best_particle $\leftarrow p$
            **else**
        **end**
        Regression_list $\leftarrow best\_particle$
    **end**
**return**

**Function** `Particle_Initialization`:
    **Input:** Regression_list, idx, dB, N_particles
    **Output:** List of the particles.
    Particles = []
    A0 = Regression_list[idx][0]
    B0 = Regression_list[idx][1]
    **for** $B$ *in* linspace(B0 − dB, B0 + dB, N_particles) **do**
        temp_particle $\leftarrow Regression\_list$
        temp_particle[idx] $\leftarrow (A0, B)$
        Particles.append(temp_particle)
    **end**
**return**

**Function** `loss`:
    **Input:** Regression_list, Idx, Experiment_data.
    **Output:** Loss_value.
    **for** (A, B) in Regression_list **do**
        OneAB_data $\leftarrow P(A, B)$
        Generated_data $\leftarrow P(\text{Regression\_list})$
        peaks = FindPeaks(OneAB_data, separation = T(A, B))
        `/* Each P(A, B) and P(Regression_list) returns generated CPMG signal`
        `   and T(A, B) returns period of P(A,B)                            */`
        Loss_value = 0
        **for** *peak in peaks* **do**
            Loss_value += L2Norm(Generated_data[peak-10:peak+10]
                        - Experiment_data[peak-10:peak+10])
        **end**
    **end**
**return**

**Supplementary Table 1. Rearranged DFT lists from the Ref 5 according to the target periods used for HPC models.** The DFT list are used to build the HPC models for identifying target spins in the strong coupling regime and for generating training datasets for all the HPC models.

|        | $W_h$ (Hz)   | $A/2\pi$ (Hz) | $B/2\pi$ (Hz) |        | $W_h$ (Hz) | $A/2\pi$ (Hz) | $B/2\pi$ (Hz) |
|--------|-------------|---------------|---------------|--------|-----------|---------------|---------------|
| $TP_{D1}$ | 138,446,962 | 137,000,000 | 19,964,000 |         | 1,233,552 | -1,189,200 | 327,800 |
|        | 138,434,443 | 137,000,000 | 19,877,000 |         | 1,221,358 | -1,176,100 | 329,400 |
|        | 138,418,692 | 137,000,000 | 19,767,000 | $TP_{D11}$ | 1,216,771 | -1,171,700 | 328,100 |
| $TP_{D2}$ | 12,431,771 | 12,374,000 | 1,197,100 |         | 1,200,629 | -1,154,900 | 328,200 |
|        | 12,400,753 | 12,342,000 | 1,205,700 |         | 1,195,339 | -1,149,200 | 328,900 |
|        | 12,370,525 | 12,313,000 | 1,191,600 |         | 1,027,859 | -890,400 | 513,500 |
|        | 12,304,586 | 12,248,000 | 1,178,700 | $TP_{D12}$ | 1,027,147 | -889,000 | 514,500 |
|        | 12,291,453 | 12,233,000 | 1,197,300 |         | 1,025,671 | -887,700 | 513,800 |
|        | 12,239,598 | 12,183,000 | 1,175,700 |         | 1,003,303 | -1,003,200 | 14,400 |
| $TP_{D2}$ | 11,322,658 | 11,225,000 | 1,483,900 | $TP_{D13}$ | 1,001,214 | -1,001,100 | 15,100 |
|        | 11,269,549 | 11,174,000 | 1,464,400 |         | 1,000,698 | -1,000,600 | 14,000 |
|        | 11,225,421 | 11,128,000 | 1,475,700 |         | 1,027,430 | 1,019,600 | 126,600 |
| $TP_{D4}$ | 8,558,610 | -8,518,900 | 823,500 | $TP_{D14}$ | 996,583 | 989,100 | 121,900 |
|        | 8,545,862 | -8,506,700 | 817,200 |         | 975,012 | 967,500 | 120,800 |
|        | 8,524,649 | -8,484,400 | 827,400 |         | 803,861 | 577,000 | 559,700 |
| $TP_{D5}$ | 6,444,101 | -6,375,600 | 937,100 | $TP_{D15}$ | 795,511 | 566,100 | 558,900 |
|        | 6,438,061 | -6,370,200 | 932,300 |         | 792,101 | 559,900 | 560,300 |
|        | 6,425,047 | -6,356,900 | 933,300 |         | 753,733 | 747,400 | 97,500 |
|        | 6,407,404 | -6,338,300 | 938,500 |         | 752,621 | 746,200 | 98,100 |
|        | 6,406,927 | -6,338,600 | 933,200 |         | 749,383 | 743,000 | 97,600 |
|        | 6,402,785 | -6,334,000 | 936,000 | $TP_{D16}$ | 744,679 | 738,400 | 96,500 |
| $TP_{D6}$ | 4,095,934 | 4,011,700 | 826,400 |         | 739,368 | 733,200 | 95,300 |
|        | 4,088,638 | 4,002,800 | 833,400 |         | 737,544 | 731,400 | 95,000 |
|        | 4,087,828 | 4,003,300 | 827,000 |         | 746,271 | 720,100 | 195,900 |
| $TP_{D7}$ | 3,697,445 | 3,621,900 | 743,600 | $TP_{D17}$ | 742,851 | 716,800 | 195,000 |
|        | 3,666,976 | 3,591,100 | 742,100 |         | 721,552 | 695,400 | 192,500 |
|        | 3,666,267 | 3,589,900 | 744,400 |         | 665,205 | 643,300 | 169,300 |
|        | 3,657,002 | 3,580,500 | 744,100 | $TP_{D18}$ | 662,828 | 641,000 | 168,700 |
|        | 3,645,855 | 3,569,800 | 740,800 |         | 661,556 | 639,500 | 169,400 |
|        | 3,638,169 | 3,561,200 | 744,400 |         | 634,404 | 603,400 | 195,900 |
| $TP_{D8}$ | 2,008,776 | 1,994,900 | 235,700 |         | 626,457 | 595,400 | 194,800 |
|        | 2,004,365 | 1,990,600 | 234,500 |         | 625,406 | 593,900 | 196,000 |
|        | 1,991,401 | 1,977,700 | 233,200 | $TP_{D19}$ | 619,048 | 587,500 | 195,100 |
|        | 1,988,363 | 1,974,700 | 232,700 |         | 596,823 | 565,300 | 191,400 |
|        | 1,978,570 | 1,964,600 | 234,700 |         | 592,530 | 560,800 | 191,300 |
|        | 1,954,927 | 1,941,100 | 232,100 |         | 540,330 | 535,100 | 75,000 |
| $TP_{D9}$ | 1,684,147 | 1,670,700 | 212,400 | $TP_{D20}$ | 529,272 | 524,200 | 73,100 |
|        | 1,670,334 | 1,656,800 | 212,200 |         | 512,548 | 507,800 | 69,600 |
|        | 1,669,221 | 1,655,200 | 215,900 |         | 470,843 | 381,900 | 275,400 |
| $TP_{D10}$ | 1,430,610 | -1,424,600 | 131,000 |         | 470,077 | 381,100 | 275,200 |
|        | 1,413,944 | -1,407,900 | 130,600 | $TP_{D21}$ | 470,018 | 381,100 | 275,100 |
|        | 1,391,684 | -1,385,400 | 132,100 |         | 468,415 | 379,700 | 274,300 |
| $TP_{D11}$ | 1,235,815 | -1,191,300 | 328,700 |         | 466,611 | 377,400 | 274,400 |
| $TP_{D22}$ | 458,262 | -228,900 | 397,000 | $TP_{D32}$ | 204,419 | -204,400 | 2,800 |
| | 457,131 | -226,100 | 397,300 | | 203,817 | -203,800 | 2,600 |
| | 454,964 | -224,000 | 396,000 | | 202,222 | -202,200 | 3,000 |
| $TP_{D23}$ | 463,963 | 373,900 | 274,700 | $TP_{D33}$ | 192,222 | -172,800 | 84,200 |
| | 416,703 | 354,700 | 218,700 | | 190,784 | -171,100 | 84,400 |
| | 416,330 | 354,200 | 218,800 | | 186,980 | -167,100 | 83,900 |
| | 414,027 | 351,800 | 218,300 | $TP_{D34}$ | 178,944 | 153,000 | 92,800 |
| $TP_{D24}$ | 376,186 | 344,900 | 150,200 | | 178,910 | 152,900 | 92,900 |
| | 372,716 | 341,200 | 150,000 | | 178,327 | 152,400 | 92,600 |
| | 370,439 | 338,400 | 150,700 | $TP_{D35}$ | 177,941 | -144,600 | 103,700 |
| $TP_{D25}$ | 310,617 | -204,500 | 233,800 | | 177,929 | -144,800 | 103,400 |
| | 309,828 | -203,300 | 233,800 | | 177,778 | -144,400 | 103,700 |
| | 309,329 | -203,000 | 233,400 | | 177,127 | -144,100 | 103,000 |
| | 309,319 | -203,100 | 233,300 | | 177,103 | -144,000 | 103,100 |
| | 308,639 | -201,600 | 233,700 | | 176,755 | -143,500 | 103,200 |
| | 308,271 | -201,500 | 233,300 | $TP_{D36}$ | 178,449 | -177,100 | 21,900 |
| $TP_{D26}$ | 298,391 | 291,000 | 66,000 | | 178,027 | -176,700 | 21,700 |
| | 295,324 | 287,900 | 65,800 | | 176,737 | -175,400 | 21,700 |
| | 294,446 | 287,000 | 65,800 | | 176,043 | -174,700 | 21,700 |
| $TP_{D27}$ | 272,905 | 224,200 | 155,600 | | 174,803 | -173,400 | 22,100 |
| | 272,002 | 223,100 | 155,600 | | 174,393 | -173,000 | 22,000 |
| | 267,791 | 219,500 | 153,400 | $TP_{D37}$ | 167,740 | 95,500 | 137,900 |
| | 266,832 | 218,400 | 153,300 | | 167,672 | 96,100 | 137,400 |
| | 265,976 | 217,000 | 153,800 | | 167,487 | 95,200 | 137,800 |
| | 265,790 | 216,700 | 153,900 | | 167,126 | 95,000 | 137,500 |
| $TP_{D28}$ | 259,997 | -193,100 | 174,100 | | 166,956 | 94,700 | 137,500 |
| | 259,239 | -192,800 | 173,300 | | 166,791 | 94,700 | 137,300 |
| | 258,956 | -192,600 | 173,100 | $TP_{D38}$ | 164,118 | -2,400 | 164,100 |
| $TP_{D29}$ | 246,006 | 243,100 | 37,700 | | 164,013 | -2,100 | 164,000 |
| | 245,732 | 242,900 | 37,200 | | 163,600 | 300 | 163,600 |
| | 245,086 | 242,200 | 37,500 | | 163,501 | -500 | 163,500 |
| | 245,002 | 242,100 | 37,600 | | 163,400 | -300 | 163,400 |
| | 243,786 | 240,900 | 37,400 | | 163,303 | -1,000 | 163,300 |
| | 243,444 | 240,600 | 37,100 | $TP_{D39}$ | 162,651 | -151,100 | 60,200 |
| $TP_{D30}$ | 214,134 | 169,000 | 131,500 | | 162,205 | -150,500 | 60,500 |
| | 213,363 | 168,100 | 131,400 | | 161,185 | -149,400 | 60,500 |
| | 211,617 | 165,800 | 131,500 | | 161,110 | -149,400 | 60,300 |
| $TP_{D31}$ | 208,668 | 97,100 | 184,700 | | 159,219 | -147,400 | 60,200 |
| | 208,563 | 96,300 | 185,000 | | 158,610 | -146,700 | 60,300 |
| | 207,385 | 94,700 | 184,500 | $TP_{D40}$ | 154,920 | 28,900 | 152,200 |
| $TP_{D32}$ | 207,115 | -207,100 | 2,500 | | 154,067 | 26,900 | 151,700 |
| | 206,516 | -206,500 | 2,600 | | 153,994 | 25,900 | 151,800 |
| | 205,416 | -205,400 | 2,600 | | 153,927 | 25,500 | 151,800 |

,

| | | | | | | | |
|---|---|---|---|---|---|---|---|
| $TP_{D40}$ | 153,152 | 23,100 | 151,400 | | 110,990 | 53,400 | 97300 |
| | 152,653 | 17,900 | 151,600 | | 110,235 | 52,000 | 97,200 |
| $TP_{D41}$ | 142,405 | 86,400 | 113,200 | $TP_{D47}$ | 110,229 | 51,800 | 97,300 |
| | 142,004 | 86,000 | 113,000 | | 110,047 | 51,600 | 97,200 |
| | 141,245 | 83,800 | 113,700 | | 109,954 | 51,400 | 97,200 |
| $TP_{D42}$ | 137,758 | -87,500 | 106,400 | | 97,219 | 78,900 | 56,800 |
| | 137,631 | -87,300 | 106,400 | | 96,998 | 78,700 | 56,700 |
| | 137,013 | -86,200 | 106,500 | | 96,894 | 78,500 | 56,800 |
| | 136,762 | -85,800 | 106,500 | | 96,858 | 78,600 | 56,600 |
| | 134,944 | -83,000 | 106,400 | $TP_{D48}$ | 96,615 | 78,300 | 56,600 |
| | 134,839 | -82,700 | 106,500 | | 94,744 | 75,300 | 57,500 |
| $TP_{D43}$ | 133,795 | -39,600 | 127800 | | 94,390 | 74,700 | 57,700 |
| | 133,736 | -39,400 | 127,800 | | 94,188 | 74,600 | 57,500 |
| | 133,647 | -39,100 | 127,800 | | 90,478 | 71,300 | 55,700 |
| | 133,436 | -37,700 | 128,000 | | 93,679 | -92,800 | 12,800 |
| | 133,358 | -38,100 | 127,800 | | 93,253 | -92,300 | 13,300 |
| | 133,081 | -37,800 | 127,600 | | 92,348 | -91,400 | 13,200 |
| $TP_{D44}$ | 128,291 | 122,900 | 36,800 | $TP_{D49}$ | 91,981 | -91,000 | 13,400 |
| | 128,234 | 122,900 | 36,600 | | 91,825 | -90,900 | 13,000 |
| | 128,167 | 122,800 | 36,700 | | 91,740 | -90,800 | 13,100 |
| $TP_{D45}$ | 119,796 | 49,700 | 109,000 | | 90,316 | -36,300 | 82,700 |
| | 119,648 | 50,000 | 108,700 | $TP_{D50}$ | 90,265 | -36,400 | 82,600 |
| | 119,622 | 49,500 | 108,900 | | 90,082 | -36,400 | 82,400 |
| | 119,614 | 49,700 | 108,800 | | 88,359 | 68,100 | 56,300 |
| | 119,557 | 50,000 | 108,600 | | 88,064 | 67,800 | 56,200 |
| | 119,531 | 49,500 | 108,800 | $TP_{D45}$ | 87,361 | 66,800 | 56,300 |
| $TP_{D46}$ | 119,496 | 100,400 | 64,800 | | 86,838 | 66,200 | 56,200 |
| | 119,130 | 99,900 | 64,900 | | 86,621 | 66,000 | 56,100 |
| | 118,706 | 98,800 | 65,800 | | 86,176 | 65,500 | 56,000 |
| | 118,552 | 99,600 | 64,300 | $TP_{D45}$ | 83,991 | 65,400 | 52,700 |
| | 118,468 | 99,500 | 64,300 | | 83,913 | 65,300 | 52,700 |
| | 118,373 | 98,400 | 65,800 | | 82,548 | 71,300 | 41,600 |
| $TP_{D47}$ | 118,475 | 114,100 | 31,900 | $TP_{D45}$ | 82,498 | 71,300 | 41,500 |
| | 116,827 | 112,500 | 31,500 | | 82,448 | 71,300 | 41,400 |
| | 116,758 | 112,400 | 31,600 | | 82,118 | -47,900 | 66,700 |
| $TP_{D46}$ | 114,455 | 73,900 | 87,400 | | 82,082 | -47,700 | 66,800 |
| | 114,274 | 73,500 | 87,500 | $TP_{D45}$ | 81,897 | -47,800 | 66,500 |
| | 114,262 | 73,600 | 87,400 | | 81,587 | -46,700 | 66,900 |
| | 114,145 | 73,300 | 87,500 | | 81,490 | -47,100 | 66,500 |
| | 114,069 | 73,300 | 87,400 | $TP_{D45}$ | 79,884 | -26,100 | 75,500 |
| | 113,992 | 73,300 | 87,300 | | 79,855 | -26,300 | 75,400 |
| $TP_{D47}$ | 111,038 | 53,500 | 97,300 | | | | |
| | 110,990 | 53,400 | 97,300 | | | | |

**Supplementary Table 2. The predicted hyperfine parameters (*A*, *B*) and corresponding confidence scores of the HPC models.** The full list of detected nuclear spins and comparison with an independent previous experiment based on multidimensional spectroscopy (a different experimental method) [3] and a previous manual analysis of DD spectroscopy (the same experimental method) [6] on the same NV center. Hyperfine parameters (*A*, *B*) calculated by DFT are denoted as $TP_{D\#}$. Confidence scores using raw (denoised) CPMG data is given. The C# with an asterisk indicates the spins identified to be strongly coupled with nuclear-nuclear interactions in Ref. [3]. C#[†] indicates a group of spins which appear as a single but broad dip in experimental CPMG data. The values (parentheses values) in the second and third columns corresponds to the confidence scores of the raw experimental data (the denoised data). Estimated errors of (*A*, *B*) values of the deep learning results are defined by standard deviation of the results with 50 iterations of fine-tuning method. Unlike in reference 3, spins with B ≅ 0 cannot be detected by the experimental sequence used in this work.

| | | | | The complete list of results | | | | | |
|---|---|---|---|---|---|---|---|---|---|
| | Target Period | HPC model confidence score N32 | HPC model confidence score N256 | Deep learning results for DD spectroscopy | | Previous multi-dimensional spectroscopy [3] | | Previous analysis for DD spectroscopy [6] | |
| | | | | $A/2\pi$ (kHz) | $B/2\pi$ (kHz) | $A/2\pi$ (kHz) | $B/2\pi$ (kHz) | $A/2\pi$ (kHz) | $B/2\pi$ (kHz) |
| C1 | $TP_{D32}$ | None | 0.98(0.99) | -213.19(5) | 4.2(9) | -213.154(1) | 3.0(4) | | |
| C2 | $TP_{524}$ | 0.95(0.98) | None | -24.34(2) | 23.2(4) | -24.399(1) | 24.81(4) | -24.4 | 26 |
| C3 | $TP_{627}$ | 0.98(0.99) | None | -20.435(0) | 39.89(2) | -20.569(1) | 41.51(3) | -20.6 | 43 |
| C4 | $TP_{645}$ | None | 0.98(0.99) | -17.69(2) | 9(1) | -17.643(1) | 8.6(2) | | |
| C5 | $TP_{708}$ | None | 0.89(0.91) | -14.61(2) | 9.7(9) | -14.548(3) | 10(1) | -14.5 | 11 |
| C6 | $TP_{842}$ | None | 0.93(0.99) | -7.81(6) | 10(1) | -7.683(4) | 4(3) | | |
| C7 | $TP_{849}$ | 0.98(0.99) | None | -8.12(1) | 22.6(7) | -8.029(1) | 21.0(4) | -8.1 | 21 |
| C8 | $TP_{897}$ | None | 0.85(0.84) | -5.02(3) | 6(2) | | | | |
| C9 | $TP_{902}$ | None | 0.87(0.88) | -4.87(6) | 6(3) | | | | |
| C10 | $TP_{923}$ | None | 0.93(0.91) | -3.78(3) | 9(1) | | | | |
| C11[†] | $TP_{936}$ ~ $TP_{950}$ | None | 0.95(0.97) | -3.07(7) | 10(2) | | | | |
| C12[†] | | None | 0.97(0.95) | -3.02(3) | 9(1) | | | | |
| C13[†] | | None | 0.92(0.96) | -2.82(6) | 8(2) | | | | |
| C14[†] | | None | 0.93(0.98) | -2.69(4) | 7(1) | -2.69(4) | 11(1) | | |
| C15[†] | | None | 0.94(0.98) | -2.67(5) | 8(2) | | | | |
| C16[†] | | None | 0.94(0.93) | -2.53(7) | 9(2) | | | | |
| C17[†] | $TP_{974}$ ~ $TP_{984}$ | None | 0.94(0.97) | -1.32(7) | 8(2) | | | | |
| C18[†] | | None | 0.97(0.99) | -1.22(2) | 7(1) | -1.212(5) | 13(1) | | |
| C19[†] | | None | 0.99(0.99) | -1.12(6) | 8(2) | | | | |
| C20[†] | | None | 0.93(0.94) | -0.99(4) | 6(2) | | | | |
| C21[†] | | None | 0.92(0.96) | -0.84(4) | 8(2) | | | | |
| C22[†] | | None | 0.95(0.95) | -0.81(6) | 8(9) | | | | |
| C23[†] | $TP_{1034}$ ~ $TP_{1044}$ | None | 0.95(0.98) | 1.711(5) | 7(2) | | | | |
| C24[†] | | None | 0.99(0.98) | 1.851(5) | 6(2) | | | | |
| C25[†] | | None | 0.99(0.97) | 2.10(6) | 8(2) | | | | |
| C26[†] | | None | 0.97(0.95) | 2.21(3) | 8(1) | | | | |
| C27[†] | | None | 0.96(0.97) | 2.25(4) | 10(2) | | | | |
| C28[†] | $TP_{1063}$ | None | 0.94(0.96) | 3.12(3) | 5(2) | | | | |
| C29[†] | ~ $TP_{1063}$ | None | 0.97(0.98) | 3.16(1) | 3.6(9) | 3.177(5) | 2(4) | | |
| - | None | None | None | None | None | 3.618(5) | 0(2) | | |
| - | None | None | None | None | None | 3.873(5) | 0(4) | | |
| C30 | $TP_{1082}$ | 0.91(0.90) | 0.88(0.91) | 4.14(6) | 8.9(5) | 4.039(5) | 0(3) | | |
| - | None | None | None | None | None | 4.225(4) | 0(6) | | |
| C31[†] | $TP_{1091}$ | None | 0.96(0.97) | 4.56(3) | 8.9(7) | | | | |
| C32[†] | ~ $TP_{1093}$ | None | 0.97(0.98) | 4.67(1) | 5.1(9) | 4.66(3) | 7(4) | | |
| C33 | $TP_{1112}$ | None | 0.95(0.99) | 5.59(6) | 9.8(5) | 5.62(1) | 5(2) | | |
| C34 | $TP_{1128}$ | None | 0.98(0.99) | 6.552(0) | 5.2(3) | | | | |
| C35 | $TP_{1147}$ | None | 0.91(0.93) | 7.40(1) | 5(1) | | | | |
| - | None | None | None | None | None | 8.32(1) | 3(5) | | |
| C36 | $TP_{1176}$ | None | 0.90(0.98) | 8.84(5) | 4.5(8) | | | | |
| C37 | $TP_{1195}$ | None | 0.95(0.97) | 9.85(1) | 8.8(5) | 9.79(2) | 5(4) | | |
| C38 | $TP_{1278}$ | None | 0.91(0.92) | 13.92(3) | 10(1) | 13.961(3) | 9(1) | | |
| C39* | $TP_{1299}$ | 0.93(0.91) | 0.95(0.97) | 14.9(8) | 12(2) | 14.07(2) | 13(1) | | |
| C40 | $TP_{1307}$ | 0.99(0.99) | None | 11.32(2) | 59.1(1) | 11.346(2) | 59.21(3) | 11.4 | 59 |
| C41[†] | $TP_{1392}$ | None | 0.93(0.93) | 19.763(0) | 7.915(0) | | | | |
| C42[†] | ~ $TP_{1395}$ | None | 0.98(0.98) | 19.79(4) | 6.1(3) | 19.815(3) | 5.3(5) | | |
| C43* | $TP_{1405}$ | 0.91(0.93) | None | 20.30(4) | 8.4(8) | 20,72(1) | 12(1) | | |
| C44* | $TP_{1463}$ | 0.83(0.85) | None | 22.9(2) | 14(2) | 23.22(1) | 13(1) | | |
| C45 | $TP_{1618}$ | None | 0.88(0.91) | 30.76(3) | 2.4(8) | | | | |
| C46* | $TP_{1634}$ | 0.92(0.93) | None | 31(1) | 7(1) | 31.25(1) | 8(2) | | |
| C47 | $TP_{1741}$ | 0.99(0.99) | None | 36.22(2) | 27.4(6) | 36.308(1) | 26.62(4) | 36.4 | 25 |
| C48 | $TP_{1972}$ | 0.88(0.91) | None | 48.55(9) | 10(2) | 48.58(2) | 9(2) | 48.7 | 12 |

**Supplementary Reference**

[1]	Robledo, L. et al. High-fidelity projective read-out of a solid-state spin quantum register. *Nature* **477**, 574-578 (2011).

[2]	Gullion, T., Baker, D. B. & Conradi, M. S. New, compensated Carr-Purcell sequences. *J. Magn. Reson.* **89**, 479–484 (1990).

[3]	Abobeih, M. H. et al. Atomic-scale imaging of a 27-nuclear-spin cluster using a quantum sensor. *Nature* **576**, 411-415 (2019).

[4]	Taminiau, T. H. et al. Detection and Control of Individual Nuclear Spins Using a Weakly Coupled Electron Spin. *Phys. Rev. Lett.* **109**, 137602 (2012).

[5]	Nizovtsev, A. P. et al. Non-flipping C-13 spins near an NV center in diamond: hyperfine and spatial characteristics by density functional theory simulation of the C-510[NV]H-252 cluster. *New J. Phys.* **20**, 023022 (2018).

[6]	Abobeih, M. H. et al. One-second coherence for a single electron spin coupled to a multi-qubit nuclear-spin. *Nat. Commun.* **9**, 2552 (2018).